\documentclass[preprint2]{aastex}
\newcommand{\be}{\begin{equation}}
\newcommand{\ee}{\end{equation}}
\newcommand{\bea}{\begin{eqnarray}}
\newcommand{\eea}{\end{eqnarray}}
\shorttitle{Detectability and Error Estimation in Orbital Fits}
\shortauthors{C.A. Giuppone et al.} 
\begin{document}
\title{Detectability and Error Estimation in Orbital Fits of Resonant Extrasolar Planets}
\author{C.A. Giuppone$^1$, M. Tadeu dos Santos$^2$, C. Beaug\'e$^1$, S. Ferraz-Mello$^2$ \\ and T.A. Michtchenko$^2$}
\affil{(1) Observatorio Astron\'omico, Universidad Nacional de C\'ordoba, C\'ordoba, Argentina}
\affil{(2) Instituto de Astronomia, Geof\'{\i}sica e Ci\^encias Atmosf\'ericas,Universidade de S\~ao Paulo, S\~ao Paulo, Brazil}

\begin{abstract}
We estimate the conditions for detectability of two planets in a 2/1 mean-motion resonance from radial velocity data, as a function of their masses, number of observations and the signal-to-noise ratio. Even for a data set of the order of 100 observations and standard deviations of the order of a few meters per second, we find that Jovian-size resonant planets are difficult to detect if the masses of the planets differ by a factor larger than $\sim 4$. This is consistent with the present population of real exosystems in the 2/1 commensurability, most of which have resonant pairs with similar minimum masses, and could indicate that many other resonant systems exist, but are presently beyond the detectability limit.

Furthermore, we analyze the error distribution in masses and orbital elements of orbital fits from synthetic data sets for resonant planets in the 2/1 commensurability. For various mass ratios and number of data points we find that the eccentricity of the outer planet is systematically over estimated, although the inner planet's eccentricity suffers a much smaller effect. If the initial conditions correspond to small amplitude oscillations around stable apsidal corotation resonances (ACR), the amplitudes estimated from the orbital fits are biased toward larger amplitudes, in accordance to results found in real resonant extrasolar systems.

\end{abstract}


\section{Introduction}

\begin{figure}[t]
\begin{center}
\includegraphics*[width=18pc]{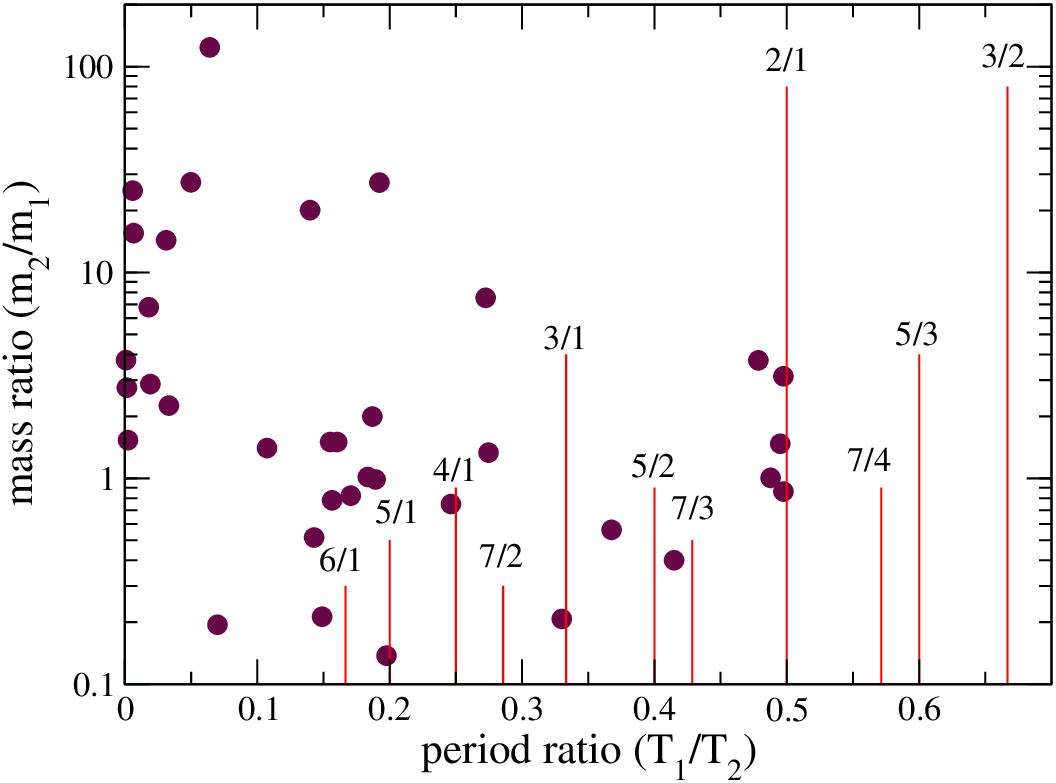}
\end{center}
\caption{Distribution of pairs of consecutive planets in multiple-planetary systems, according to mass ratio over ratio of orbital periods. MMRs are indicated by the vertical red lines, whose length is inversely proportional to the order of the MMR. The data correspond to most current orbital fits.}
\label{fig1}
\end{figure}

Although the population of extrasolar planets continues to increase rapidly, the number of multiple-planet systems with members in mean-motion resonances (MMRs) shows a much slower growth rate. The existence of a 3/1 commensurability in 55Cnc has been recently questioned by the five planet fit of Fischer et al. (2008) whose best solution now indicates a non-resonant motion for 55Cnc-c and 55Cnc-d. Even in the 2/1 MMR, the most dynamically important and populated commensurability, the number of confirmed planetary systems is still restricted to four well known cases: GJ876, HD82943, HD73526 and HD128311. Go\'zdziewski et al. (2007) proposed that HD160691 may also have two planets in the 2/1 MMR, although this is unconfirmed, and just two years before the best fit solution seemed to favor a 5/1 MMR between the same two planets (Go\'zdziewski et al. 2005). Fig.~\ref{fig1} shows the distribution of exoplanet pairs according to their mass ratio and orbital period ratio. Except for the vicinity of the 2/1, they appear more or less at random. 

However, this picture may change in the near future. Recently, two possible new resonant systems have been proposed based on radial velocity RV observations from the Geneva group. Laskar and Correia (2009) have found a planetary system around HD60532 which appears to be trapped in a 3/1 MMR, while Correia et al. (2009) show a similar behavior for the orbital solutions of a pair of planets around HD45364, this time in around a 3/2 commensurability. Given the volatility of orbital fits for near-resonant configurations (e.g. 55Cnc) it is perhaps too early to treat these cases as confirmed. Nevertheless, their importance is unquestionable, specially since they would populate commensurabilities which are presently empty.

For several years a large proportion of resonant planets was expected as a consequence of an assumed large-scale planetary migration of exoplanets due to interactions with the gaseous disk. In fact, the existence of resonant systems has many times been advanced as observational evidence that such a migration actually took place, and many planets were formed farther from the star than their present location. Hydrodynamical simulations seem to indicate that resonance trapping in low-order commensurabilities (particularly the 2/1 MMR) are high-probability outcomes for a wide spectrum of planetary masses, initial conditions and disk parameters. Several works, particularly focused on GJ876 (see, e.g., Kley et al. 2005), point that the present configuration of planets b and c can only be explained assuming such a scenario. 

Three explanations have been presented recently to account for the lack of a larger resonant population. One possibility is that not all planets approaching the 2/1 resonance may have been captured. If the mass ratio between the outer and inner planet was sufficiently small (of the order of the Saturn over Jupiter ratio), then a very fast orbital decay of the outer smaller body (e.g., Type III migration) may have avoided resonance capture in the 2/1 and led to a later trapping in the 3/2 (Masset \& Snellgrove 2001). A similar effect has been proposed by Morbidelli \& Crida (2007) as a first step to explain the current orbital architecture of the outer planets of our own solar system. However, it is not clear under what circumstances such a fast migration would occur (see D'Angelo \& Lubow 2008). Alternatively, turbulence effects in the gaseous disk may have caused significant orbital perturbations in the decaying bodies to inhibit resonance trapping (Adams et al. 2008). 

A second possibility may lie in the survival rate of resonant planets during their evolution within the commensurability. As shown originally by Lee and Peale (2002), once inside the resonance domain, tidal interactions with a gaseous disk will increase the eccentricities of the planetary bodies until a disruptive collision or ejection of one body occurs. This is due to the topology of the stable ACR families in the eccentricity domain (see, e.g., Beaug\'e et al. 2006, Hadjidemetriou 2008) within the 2/1 MMR. The only way two resonant planets may survive a large scale orbital migration is if the driving mechanism introduced a significant damping of the orbital eccentricities, leading to equilibrium values of these elements comparable with the observed values (Kley et al. 2005, Beaug\'e et al. 2006). Although this effect is expected from nominal disc parameters, especially if an inner disk is assumed (Crida et al. 2008), there is no evidence that this must be true in all cases, and perhaps most of the systems originally trapped in the 2/1 could have been ejected. Moorhead and Adams (2005) proposed such a mechanism to explain the present semimajor axis and eccentricity distribution of exoplanets. 

A third possibility, little considered up to now, is that the apparent lack of resonant planets may be due to detectability limitations. Recall that resonant motion causes an almost periodic repetition of the RV curve of the star which, under certain circumstances, may not allow a good separation of components. Perhaps, many additional systems may actually lie within the 2/1 resonance but are currently not discernible due to limited observations or small signal-to-noise ratios. Recently, Anglada-Escud\'e et al. (2008) discussed cases where a two-planet resonant system with almost circular orbits may appear masked as a single planet in an elliptical orbit of eccentricity $e$. However, this effect appears to be possible only in cases where the outer resonant planet is much more massive than its inner companion and the single-planet solution has a low eccentricity (e.g., $e < 0.2$). In this paper we address this question in more detail constructing a detectability criterion valid for any mass ratio and not restricted to quasi-circular orbits.

We also discuss the errors in the planetary masses and orbital parameters of those resonant systems which can actually be detected. Although this is different problem, it shares many common points and can be studied using the same approach. Of the four systems currently inhabiting the 2/1 MMR, only GJ876 has a dynamically stable best fit, while the others are characterized by orbits leading to a disruption of the system in time scales much smaller than the age of the star. Although stable orbital solutions are possible for these troublesome cases, having rms similar to the best fit, they usually correspond to large amplitude ACR, not easily compatible with a smooth orbital migration (S\'andor et al. 2007, Crida et al. 2008).

\section{Orbital Fits with Synthetic Data Sets}

In order to avoid the problems of estimating errors in observational data with unknown solution, we will work with synthetic data sets of RVs. We will assume the existence of two planets of masses $m_1$ and $m_2$ orbiting a star $m_0$ with semimajor axes $a_i$, eccentricities $e_i$, mean longitudes $\lambda_i$ and longitudes of pericenter $\varpi_i$. The index $i=1$ will be used for the inner planet, while $i=2$ will denote the outer body (i.e. $a_1 < a_2$). We assume that both bodies share the same orbital plane oriented edge-on with respect to the observer.

From the nominal solution, we generate a synthetic RV curve mimicking the star movement around the barycenter of the system. This curve is the sum of two periodic signals, each with amplitude $K_j$ related to the $j$-th planet. Our RV data set will only cover a few orbital periods, and we will assume no significant effects from mutual gravitational perturbations.

Once the synthetic curve is generated, we construct a discrete sampling choosing $N$ observation times $t_i$ distributed randomly according to a homogeneous distribution. At each point, we calculate a RV value as a random displacement of the nominal $V_r(t_i)$; this displacement follows a Gaussian distribution with constant variance $\sigma^2$. The resulting synthetic data set will consist of three columns $(t_i,{V_r}_i,\sigma)$ and will be used as input in our orbital fitting procedure (Beaug\'e et al. 2008).

\begin{table}[t]
\begin{center}
\begin{tabular}{|c|c|c|c|c|c|c|c}
\hline 
Planet & Mass & $a$ & $e$  & $M$ & $\varpi$ \\
\hline 
 1 &  1.0   &    0.6298   &  0.4137   &    0.0   &   0.0   \\
 2 &  3.0   &    1.0   &  0.0946   &    0.0   &   0.0   \\
\hline
 1 &  1.0   &    0.6299   &  0.3923   &    0.0   &   0.0   \\
 2 &  1.0   &    1.0   &  0.1271   &    0.0   &   0.0   \\
\hline 
 1 &  1.0   &    0.6299   &  0.1105   &   310.4  &   115.9   \\
 2 &  0.33   &   1.0031   &  0.3840   &    65.30  &   0.0   \\
\hline 
\end{tabular} 
\end{center}
\caption{Orbital Parameter for Three ACR Solutions in the 2/1 MMR for Different Mass Ratios $m_2/m_1$. Masses are in units of Jovian mass and angles in degrees. The central mass is $m_0 = 1 M_{\odot}$.}
\label{table1}
\end{table}

\begin{figure}[t]
\begin{center}
\includegraphics*[width=18pc]{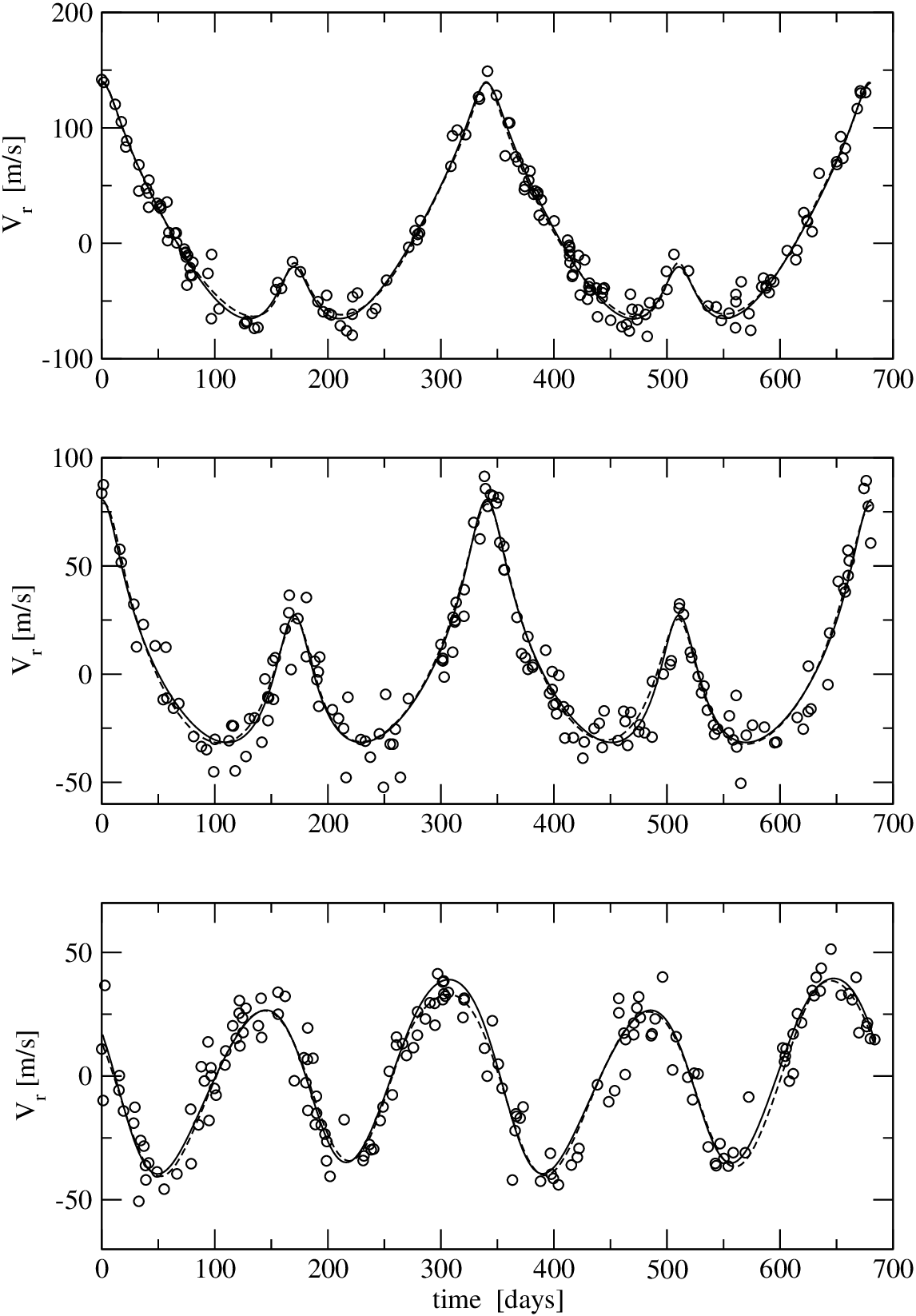}
\end{center}
\caption{Black continuous line shows synthetic RV curves for the three ACR presented in Table \ref{table1}. Mass ratio decreases from top to bottom. Open circles are $N=165$ fictitious points with random time distribution and standard deviation $\sigma=10$ m s$^{-1}$.}
\label{fig2}
\end{figure}

\subsection{Two-Planet Systems in a 2/1 Mean-Motion Resonance}\label{two}

Consider two planets in a small-amplitude oscillation around a stable ACR in the 2/1 MMR. The question is how the uncertainties in the orbital fit will affect the observed motion of the system. As examples, we have chosen three different configurations, each corresponding to a stable ACR with different mass ratios $m_2/m_1$. The first two correspond to ACR of type $(0,0)$, while the last displays an asymmetric corotational behavior. Masses and initial orbital elements are given in Table \ref{table1}. In all cases the initial conditions lead to a small amplitude oscillation (approximately 5 degrees) around the respective ACR. 

Typical examples of synthetic RV data sets with $\sigma=10$ m s$^{-1}$ are shown in Fig.~\ref{fig2}. Although all the generated data sets are constructed with the same standard deviation $\sigma$, the signal-to-noise ratio appears different for each mass ratio. To understand this effect, let us write the total amplitude of the RV curve as ${\rm amp}(V_r) = K_1 + K_2$. Expressing $K_i$ in terms of the mass and orbital elements (see, e.g., Beaug\'e et al. 2007) for an edge-on coplanar system, we can write (up to second order in the masses, and for quasi-circular orbits):
\be
\label{eq12}
{\rm amp}(V_r) = K_1 + K_2 = {m_1 \over m_0} n_1 a_1 + {m_2 \over m_0} n_2 a_2,
\ee
where $n_i$ are the mean motions. For planets in the vicinity of a 2/1 MMR, we can simplify this expression as:
\be
\label{eq13}
{\rm amp}(V_r) \simeq \biggl( 1 + \sqrt[3]{{1 \over 2}}{m_2 \over m_1}                         \biggr)  {m_1 \over m_0} n_1 a_1,
\ee
where $(n_2/n_1)(a_2/a_1) \simeq \sqrt[3]{1/2}$. The quantity inside the brackets represents the increase in the RV amplitude due to the presence of the outer (resonant) planet. This term tends to unity for $m_2 \rightarrow 0$, and shows a linear dependence with the mass ratio. Thus, larger values of $m_2/m_1$ will produce a larger RV signal and, assuming a fixed observational standard deviation $\sigma$, will result in a larger signal-to-noise ratio.

Equation (\ref{eq13}) is also a rough indication of which planet dominates the RV signal. The critical mass ratio is given by $m_2/m_1 \simeq \sqrt[3]{2} \simeq 1.26$. For smaller values, the RV amplitude of the inner planet $m_1$ is larger, and the RV curve appears as a perturbed signal with primary period equal to the orbital period of the inner planet (i.e., $2 \pi/n_1$). One example is shown in the bottom frame of Fig.~\ref{fig2}. Conversely, if the mass ratio is larger than $\simeq 1.26$, the signal from the outer planet becomes larger and the dominant period in the RV curve is given by $2 \pi/n_2$ (see top graph of Fig.~\ref{fig2}). The middle frame represents a transition region in which both components are of similar magnitude.

\section{Detectability Criteria}

Cumming (2004) presented a simple procedure to estimate the detectability of single-planetary systems, given the number of data points $N$ and $K/\sigma$ ratio, as a function of the desired false-alarm probability (FAP). Although the original formulation was developed for circular orbits and for large data sets, it serves as a starting point for the extensions shown below. Throughout this paper we will assume that the observational time frame covers at least one orbital period of the planetary masses.

Imagine that we have two fits of a given RV data set ($V_i$) with $N$ points. Each fit is assumed to contain $\nu=N-M$ degrees of freedom (i.e., $M$ free parameters) and to yield a squared sum of residuals
\be
\label{eqs1}
Q = \sum_1^N (V_i - W(t_i))^2
\ee
where $W(t)$ is the adopted model\footnote{We prefer to use $Q$ instead of $\chi^2$ $(=Q/\sigma^2)$ because of the ambiguity introduced on this notation by its improper usage in astronomy where $\chi^2$ is often ''normalized" and not the same quantity used in statistics.}. As usual, the $V_i$ are assumed to be statistically independent normal variates $V_i=N(W(t_i),\sigma^2)$. The variance $\sigma^2$ (unknown) is time-independent but, if necessary, the definition of $Q$ can be changed to introduce weights and take into account different variances (see Ferraz-Mello, 1981). 

We adopt the subscripts $a$ and $b$ to the first and second fit, respectively. In addition, we assume that the model $a$ is embedded into model $b$ ($M_a<M_b$). The improvement of the goodness of fit may be characterized by the power $z$ defined as (see Cumming 2004)
\be
z=\frac{\nu_b}{\nu_a-\nu_b}\frac{Q_a-Q_b}{Q_b} .
\label{z-floating}
\ee
This formula extends to the general case the "floating-mean power" introduced in the study of periodograms by Cumming et al. (1999). When dealing with a Gaussian white noise, the probability distribution functions of this statistic is given by the Fisher-Snedecor distribution $F_{\nu_a-\nu_b,\nu_b}(z)$. 

The FAP (or probability of error) is the probability of getting one result just by chance when working with a white noise. When the result is the solution of the minimum problem $Q=Q_{min}$, we must keep in mind that it does not refer to a random choice of the parameters, but to the maximum $z$ obtained from a given number of trials. Following Cumming (2004), the FAP ${\cal F}(\widehat{z})$ of the best fit result $\widehat{z}$ is given by
\be
{\cal F}(\widehat{z}) = {\cal M} \; \int_{\widehat{z}}^\infty F_{\nu_a-\nu_b,\nu_b}(z) dz,
\label{probzd}
\ee
where ${\cal M}$ is roughly the number of independent trials. In the case of a periodogram constructed with equidistant data points, the periodogram is a Fourier transform and ${\cal M}=N/2$. For more involved problems, there is no simple expression and this quantity is often determined from Monte Carlo simulations. When the data are unequally spaced, empirical formulae are used. We mention the rules introduced by Cumming (2004) and by Quast (Ferraz-Mello and Quast, 1987). In the cases studied in this paper, we have found that ${\cal M} \sim N$ seems to give a better agreement with Monte Carlo simulations. This is the value adopted through the present work. Equation (\ref{probzd}) can then be inverted to give the detectability limit $z_d$ corresponding to a user specified ${\rm FAP}_d$, i.e., solving 
\be
{\rm FAP}_d={\cal F}(z_d)
\label{fapd}
\ee
to obtain the necessary $z_d$. 

\subsection{Synthetic random samples}\label{sec3.1}

In order to study how the detectability varies with the parameters and the quality of the available measurements, we may study a simple model $b$ in which we assume that the measurements are given by $V_i = \mathcal{W}_b(t_i) + \epsilon_i$ where $\mathcal{W}_b(t)$ is a chosen model (circular, Keplerian with $e\neq 0$, etc.) and $\epsilon_i$ is a Gaussian noise (normal variate) $N(0,\sigma^2_\epsilon)$ uncorrelated with $\mathcal{W}_b$. As model $a$, we use the mean of the data, assumed to have been averaged to zero beforehand. We may use orthonormal functions (see the Appendix) to get an approximation for the statistic ${z}$ corresponding to the best-fit solutions\footnote{Throughout the paper the symbol $\,\widehat{}\,$ over a given quantity indicates its best-fit estimation.}. We can rewrite equation (\ref{zduplo}) as
\be
\label{z-modelo}
\widehat{z} = \frac{\nu_b}{\nu_a-\nu_b}\left(\frac
{||\mathcal{W}_b||+(N-1)\sigma^2_\epsilon}{\widehat{Q}_b} - 1\right) ,
\ee
which may be simplified to read
\be
\widehat{z}_1 = \frac{\Lambda\sigma^2_\epsilon}{\widehat{Q}_b}
\label{z1}
\ee
where we have introduced the auxiliary statistic 
\be
\widehat{z}_1=\widehat{z} +\frac{\nu_b}{\nu_a-\nu_b}
\ee
and the constant 
\be
\Lambda=\frac{\nu_b}{\nu_a-\nu_b}\left(\frac{N \langle\mathcal{W}_b^2\rangle}{\sigma^2_\epsilon}+N-1\right).
\ee
In the case of one planet in circular orbit and homogeneously distributed observations, $\langle\mathcal{W}_b^2\rangle=K^2/2$ (see the Appendix and Cumming 2004).

\begin{figure}[t]
\begin{center}
\includegraphics*[width=18pc]{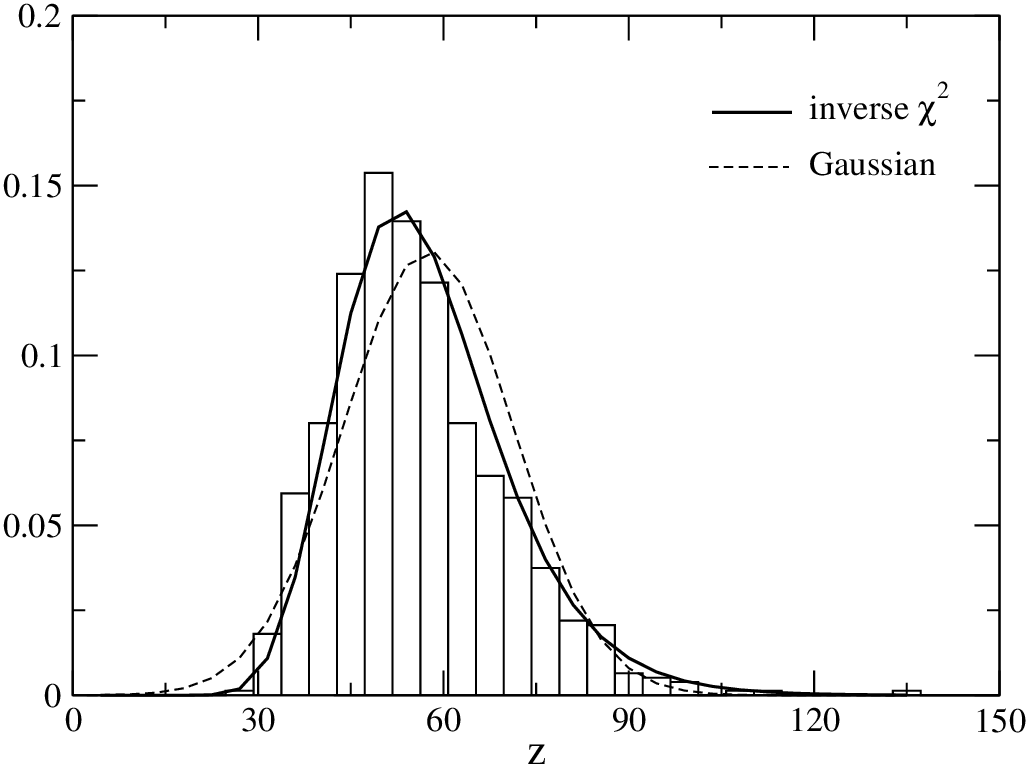}
\end{center}
\caption{Histogram of the best-fit values $\widehat{z}$ resulting from 774 random data sets constructed for a single planet ($m_1=1 M_{\rm Jup}$) in a circular orbit with orbital period of $100$ days. For each sample the initial mean longitude was chosen randomly, and the corresponding RV curve was sampled with $N=60$ and $K/\sigma = 3$. The solid curve shows the inverse chi-square distribution using equation (\ref{prob}), while the dashed line corresponds to the Gaussian distribution with same mean and variance. All plots subtend a unit area.}
\label{fig3}
\end{figure}

The only random quantity in equation (\ref{z1}) is $\widehat{Q}_b$. It is known from linear regression theory (see Kurth, 1967) that  $\widehat{Q}_b/\sigma^2_\epsilon$ is a $\chi^2_\nu$ variate with $\nu=\nu_b$. Therefore, the statistic $z_1$ is an inverse chi-square distribution, and 
\be
\label{prob}
{\rm Prob}(z>z_d) = {\rm Prob}\left(z_1>z_{1d}\right)
      = {\rm Prob}\left(\frac{\Lambda}{\chi^2_\nu} > z_{1d}\right) .
\ee
From these expressions, we can obtain the mean and variance of $\widehat{z}$ as
\bea
\label{z-medio}
\langle\widehat{z}\rangle &=& E(\widehat{z}) =
  \frac{\Lambda}{\nu_b-2} -\frac{\nu_b}{\nu_a-\nu_b} \nonumber \\
&&  \\
\sigma_{\widehat{z}}^2 &=& D^2(\widehat{z}) =
        \frac{2 \Lambda^2}{(\nu_b-2)^2(\nu_b-4)} \nonumber .
\eea
In the approximation for large $N-M_b$, the standard deviation of $\widehat{z}$ is approximately given by
\be
\label{sigma-z}
\sigma_{\widehat{z}} \simeq \frac{\sqrt{2}N}{(M_b-1)\sqrt{N-M_b}}\left(\frac{\langle \mathcal{W}_b^2 \rangle}{\sigma^2_\epsilon} + 1\right).
\ee

These results may be compared with those published by Cumming (2004). If we adopt the approximations $\langle\mathcal{W}_b^2\rangle=K^2/2$
and $\sigma_\epsilon = \sigma$, the results for $\langle\widehat{z}\rangle$ are not very different from those given above and tend to be equal for very large $N$. However, Cumming's results for $\sigma_{\widehat{z}}^2$ are significantly smaller than those given by equation (\ref{z-medio}) (e.g., $9.8$ instead of $13.7$ in the example considered in Fig.~\ref{fig3}). The corresponding Gaussian is then much more narrow than than shown in Fig.~\ref{fig3} and does not reproduce satisfactorily the distribution of the values of $\widehat{z}$. 

For a given signal amplitude $K$, the distribution of $\widehat{z}$ gives the dispersion of possible powers obtainable from different data sets with the same $N$ and signal-to-noise ratio. Thus, for given detection threshold $z_d$ (or $z_{1d}$), we can look for the value of the parameters corresponding to a given probability of detection $P_{\rm detect}$. This is done solving the equation ${\rm Prob}\left(\frac{\Lambda}{\chi^2_\nu} > z_{1d}\right)= P_{\rm detect}$
with respect to $z_{1d}$. Adopting a numerical value for $P_{\rm detect}$ (e.g., $99\%$), equations (\ref{prob})-(\ref{z-medio}) yield the necessary value of $\Lambda$. Finally, from the  approximation  $\langle\mathcal{W}^2\rangle/\sigma^2_\epsilon=K^2/(2\sigma^2_\epsilon)$ we can obtain a relation between $K/\sigma_\epsilon$ and $N$ to satisfy this criterion. It is worth mentioning that the results thus obtained are very robust and do not depend critically on the chosen distribution of the best-fit values $\widehat{z}$.

\subsection{Detectability of Two Resonant Planets}

To estimate the detectability limit for two planets in the 2/1 MMR, we assume that the body generating the largest amplitude in the RV signal is already known, and wish to calculate under which conditions the signal of the second body also satisfies the given detectability limit. A look at the different plots of Fig.~\ref{fig2} shows that there is no unique separation of the signals. When the mass ratio $m_2/m_1$ is large, the RV amplitude of the outer planet is also large (implying $K_2 > K_1$), and the detectability of the resonant pair reduces to the discernability of the inner mass. In the opposite case where $K_1 > K_2$, it is the inner planet that dominates the RV curve and the existence of the outer body must be deduced from the difference between successive maxima. 

Extending the equations in Section \ref{two} to elliptic orbits, we can write the ratio of amplitudes of both planets as
\be
\label{eq13bis}
{K_2 \over K_1} = \sqrt[3]{{1 \over 2}} \; { m_2 \over m_1} \sqrt{{1 - e_1^2 \over 1 - e_2^2}} 
\ee
showing that the value $m_2/m_1 \sim \sqrt[3]{2}$ deduced earlier still provides a qualitative critical value for $K_1 \sim K_2$ when both planets have similar eccentricities.

We begin discussing the case of a more massive inner planet. Here $K_1 > K_2$, and we wish to estimate the detectability limit of finding $m_2$ assuming that $m_1$ is known. We can employ exactly the same criterion constructed for one-planet systems, where the number of free parameters are now $M_b=11$ and $M_a=6$, and rewrite (\ref{zfin}) as
\be
\label{eq14}
\widehat{z} = \frac{\nu_b}{\nu_a-\nu_b}\left(\frac
{||\mathcal{W}_2||+(N-1)\sigma'^2_\epsilon}{\widehat{Q}_b} - 1\right) .
\ee
If we assume $\sigma'_\epsilon \simeq \sigma$ (see the Appendix), the remaining procedure is analogous to that deduced in Section \ref{sec3.1}. It is then possible to estimate a limiting value of $K_2/\sigma$, as function of the number of points $N$ necessary for the resulting peak to surpass a pre-established FAP value with a probability equal to $P_{detect}$. Moreover, since 
\be
\label{eq20}
{K_2 \over \sigma} = \biggl( {K_2 \over K_1} \biggr) 
                       \biggl( {K_1 \over \sigma} \biggr)
\ee
we can also address this problem saying that the detectability condition depends on $N$ and on the ratios $K_1/\sigma$ and $K_2/K_1$. Applying equation (\ref{eq13bis}) we can transform the condition in $K_2/K_1$ to one in $m_2/m_1$. Since $K_2<K_1$, the question reduces to finding what values of $m_2/m_1$ (for given eccentricties $e_i$) reduce the value ${K_2/\sigma}$ to undetectable magnitudes. 

The ratio ${K_1/\sigma}$ depends on the standard deviation of the data, the stellar mass, as well as on the orbital parameters and mass of the inner planet. For example, larger values of $m_1$ will increase the amplitude of $K_1$, thus allowing the detection of smaller values of $m_2/m_1$ for a fixed $\sigma$. 
Let us call ${K_1}_0$ the RV amplitude generated by a Jovian size planet ($m_1 = M_{\rm Jup}$) at $a_1=1$ AU and eccentricity $e_1$. Since $K_1 \propto n_1 a_1 (m_1/m_0)$, for any other mass and semimajor axis, the ratio ${K_1/\sigma}$ may be 
written as
\be
\label{eq20tris}
{K_1 \over \sigma} = {{K_1}_0 \over \sigma^*},
\ee
where the ``scaled'' standard deviation $\sigma^*$ is related to the nominal value $\sigma$ by
\be
\label{eq20cuatris}
\sigma = \sigma^* \biggl( {m_1 \over M_{\rm Jup}} \biggr) 
                  \biggl( {a_1 \over 1 {\rm AU}} \biggr)^{-1/2}
                  \biggl( {m_0 \over M_{\odot}} \biggr)^{-1}.
\ee
Finally, from equations (\ref{eq13bis}),(\ref{eq20}) and (\ref{eq20tris}) we can then write
\be
\label{eq20bis}
{K_2 \over \sigma} = \sqrt[3]{{1 \over 2}} \; \sqrt{{1 - e_1^2 \over 1 - e_2^2}} 
\; { m_2 \over m_1} \biggl( {{K_1}_0 \over \sigma^*} \biggr) .
\ee
This equation allows us to estimate the minimum value of $m_2/m_1$ for detectability of the outer resonant planet as function of $N$ and $\sigma^*$, for given orbital eccentricities $e_i$. If we assume that the inner planet corresponds to a Jovian mass at $1$ AU, $\sigma^*$ is equal to the nominal standard deviation
of the RV data $\sigma$. For other inner masses, $\sigma$ can be calculated through
(\ref{eq20cuatris}).

\begin{figure}[t]
\begin{center}
\includegraphics*[width=18pc]{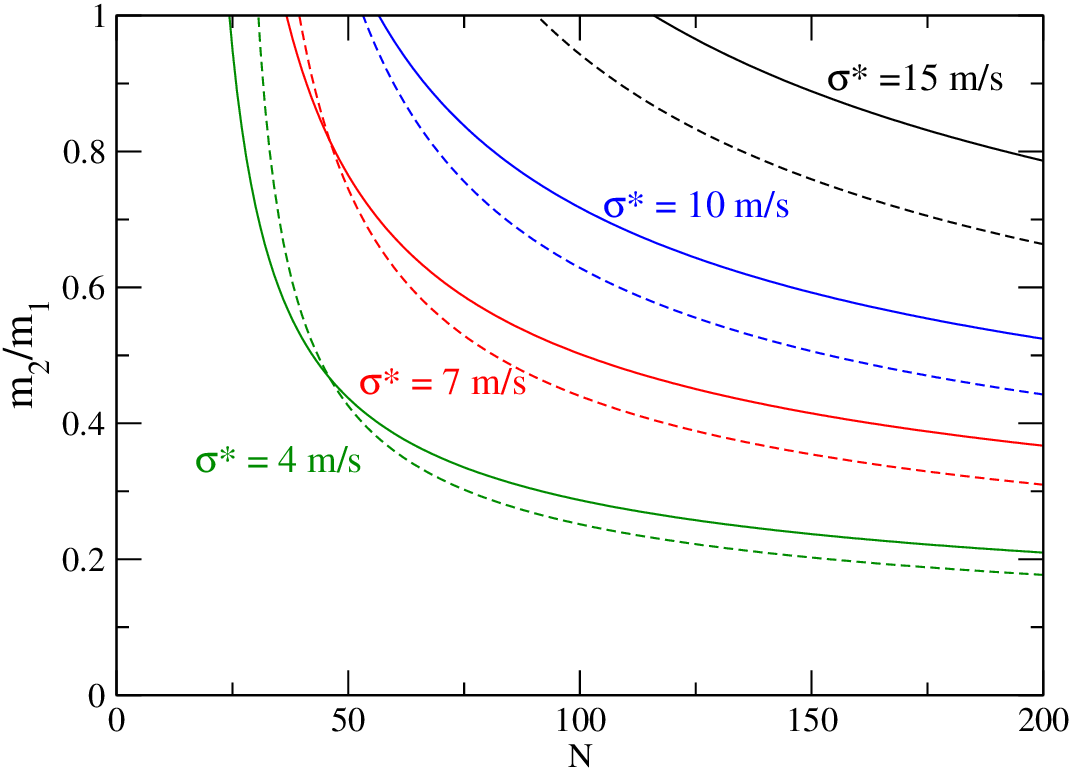}
\end{center}
\caption{Minimum value of mass ratio $m_2/m_1$ necessary for  detectability of the smaller outer resonant planet (with $P_{detect}= 0.99$), as a function of number $N$ of data points and for four values of the scaled standard deviation $\sigma^*$. Both planetary eccentricities are chosen as $e_i = 0.1$. The FAP is taken equal to ${\cal F} = 10^{-4}$. The continuous lines correspond to predictions with the inverse chi-square distribution (\ref{prob}), while dashed curves are obtained with the Gaussian approximation with equal mean and variance.}
\label{fig4}
\end{figure}

Fig.~\ref{fig4} shows the detectability limit for $m_2/m_1 < 1$ for several values of $\sigma^*$ and as a function of the number $N$ of data points. Each curve gives the minimum value of the mass ratio $m_2/m_1$ necessary to achieve a FAP of ${\cal F}=10^{-4}$. The stellar mass was $m_0 = 1 M_{\odot}$ and both eccentricities were chosen equal to $0.1$. Continuous lines were constructed using the inverse chi-square distribution (\ref{prob}) while the dashed curves denote approximate values obtained with a Gaussian distribution with same mean and variance. Qualitatively both models yield to similar results, although expression (\ref{prob}) leads to more restrictive detectability limits.

\begin{figure}[t]
\begin{center}
\includegraphics*[width=18pc]{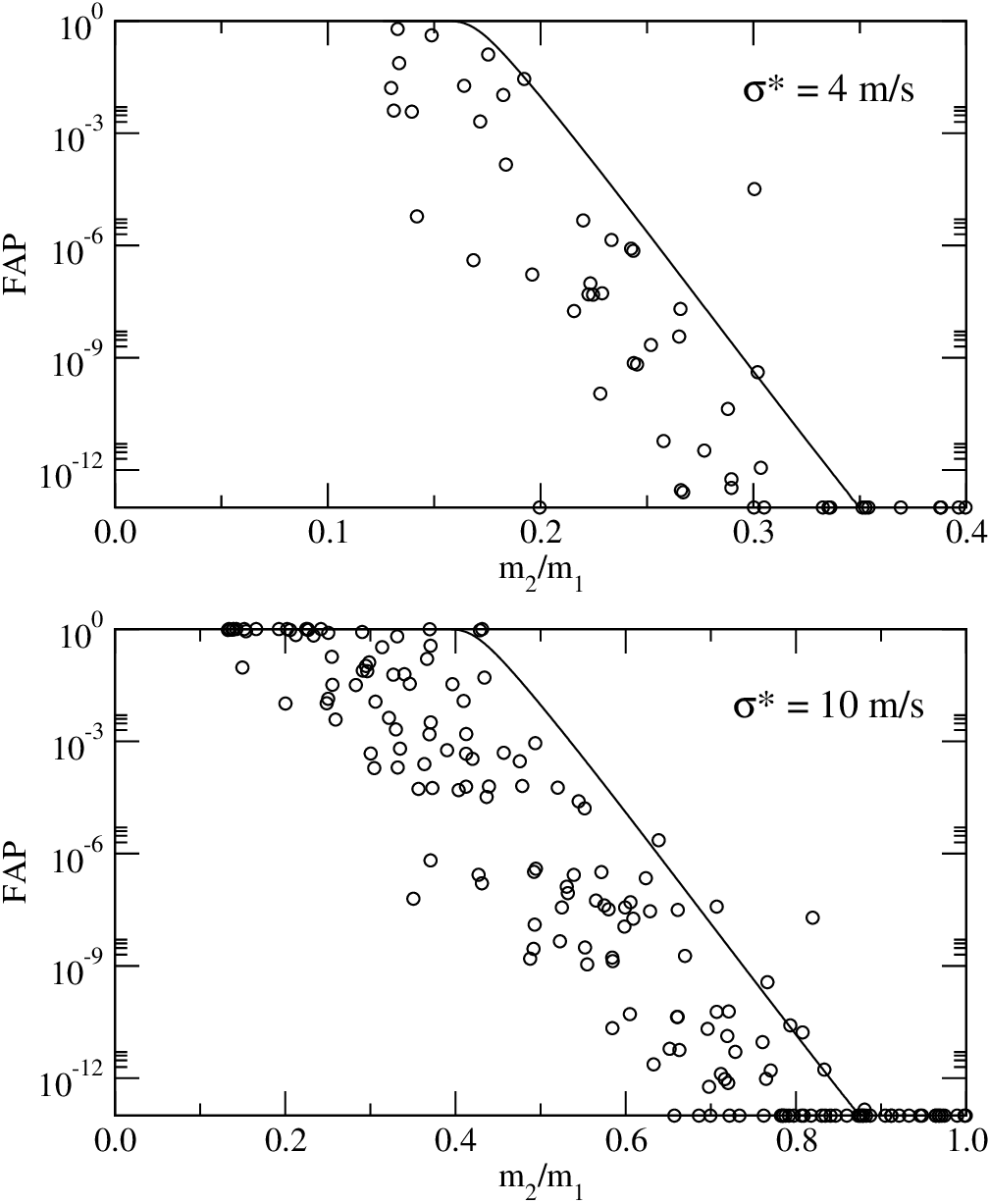}
\end{center}
\caption{Detectability of two planets in the 2/1 MMR. Open circles show results of Monte Carlo simulations from synthetics data sets. The plot show the FAP for random values of the mass ratio $m_2/m_1$. The total number of points was $N=100$ distributed randomly over four orbital periods of the outer body. The broad continuous curves show prediction values from the model with $P_{detect}= 0.99$.}
\label{fig5}
\end{figure}

For $\sigma^* = 15$ m s$^{-1}$, practically no resonant system is detectable with mass ratio $m_2/m_1$ lower than $\sim 0.6$, even for a data set containing $N=200$ RV observations. For $N < 60$ practically no system with this $\sigma^*$ is detectable for any mass ratio. The picture improves with lower values of the scaled standard deviation, until for $\sigma^* = 4$ m s$^{-1}$ it is possible to detect systems with mass ratio close to one-tenth provided the data set is sufficiently large. 

It is important to note that the value of $\sigma$ to consider in these calculations is not just the one deduced from the observational techniques, but the total standard deviation including stellar jitter and the possible incompleteness of the two-planet model. Typical values of stellar jitter (Wright 2005) are of the order of $4-7$ m s$^{-1}$, and instrumental uncertainties are also of similar magnitude. The square sum of both implies that typical values of $\sigma$ for these systems should be of the order of $6-10$ m s$^{-1}$. A good qualitative estimate of $\sigma$ may be given by the value of the rms resulting from the orbital fit (Shen \& Turner 2008) which leads to $\sigma \sim 7-10$ for most multiple-planet systems. 

To check the precision of the limits obtained with this model, we performed a series of Monte Carlo simulations with synthetic data sets, each with $N=100$ data points covering a total of four orbital periods of the outer resonant body and with mass ratios in the interval $m_2/m_1 \in [0.1,1.0]$. The inner planet was fixed to the same mass and orbital parameters as used in Fig.~\ref{fig4} ($m_1 = 1 M_{\rm Jup}$, $a_1=1$ AU). One series of simulations was done adopting $\sigma^*=4$ m s$^{-1}$, while a second run considered $\sigma=10^*$ m s$^{-1}$. For each data set we determined the best two-planet fit and calculated the value of the FAP corresponding to the second planet using the same procedure as for the one-planet system. Results are shown in Fig.~\ref{fig5}, where the open circles show the relationship between mass ratio and the respective value of FAP for all the data sets. Values of the ordinate lower than $10^{-13}$ were equated to this lower limit. Finally, the solid curves present the analytical detectability limits obtained from the model. 

These plots must be read in the following way. For a given value of the mass ratio $m_2/m_1$, all the synthetic data sets give orbital fits with FAP lower than the broad continuous curve. Thus, in order to guarantee detectability, the FAP corresponding to a given mass ratio must be above that threshold. The agreement between the analytical curve and the numerical simulations is very good, indicating that the simplified model described above is more than adequate to predict the detectability of multiple planet system.

\begin{figure}[t]
\begin{center}
\includegraphics*[width=18pc]{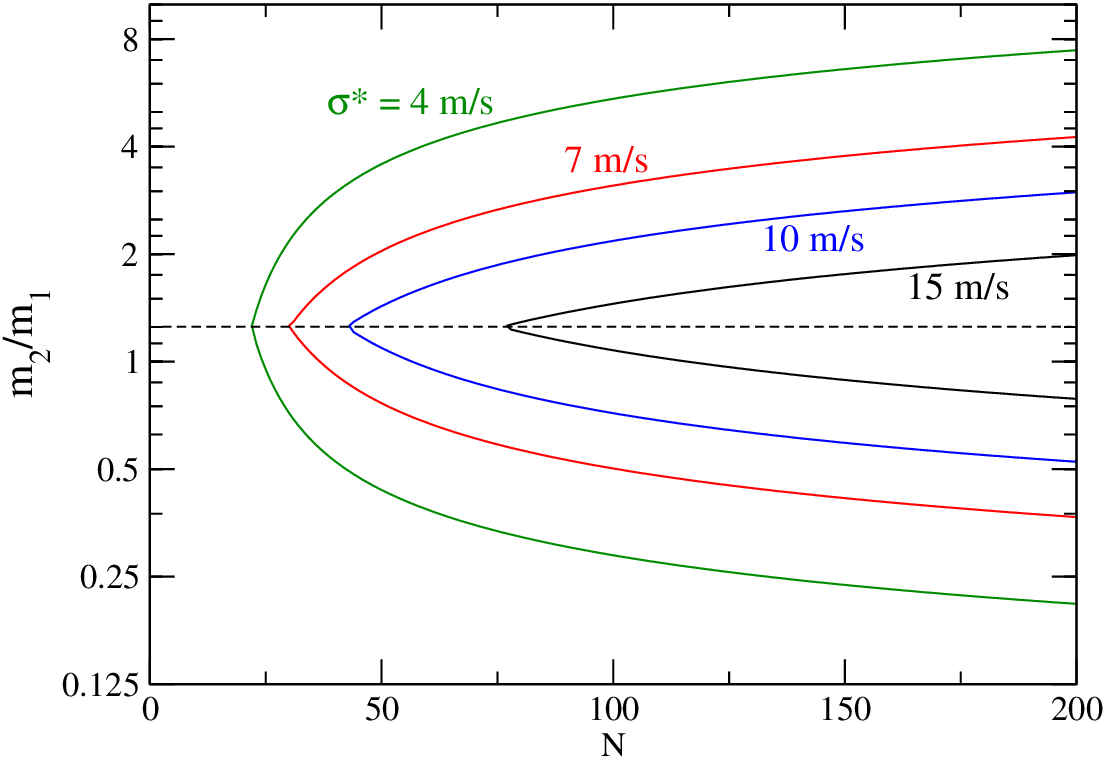}
\end{center}
\caption{Same as Fig.~\ref{fig4}, but including mass ratios $m_2/m_1>1$.
Only results constructed with the inverse chi-square distribution (\ref{prob}) are shown. For a given value of $\sigma^*$, detectability is guaranteed only inside the two corresponding curves.}
\label{fig6}
\end{figure}

The same calculations can be done for systems with $K_2/K_1 > 1$. In such a case the detectability criterion follows the same route, but now the outer planet is responsible for the dominant peak, while the signal that defines the detectability of the resonant system is $K_1$. Consequently, equation (\ref{eq20}) must now be inverted and will yield the minimum necessary value of $K_1/K_2$ such that the value ${K_1/\sigma}$ is still detectable. The dependence with the individual masses and semimajor axis can again be overcome defining the ``scaled'' standard deviation $\sigma^*$ as in equation (\ref{eq20cuatris}). Note that we have kept the mass and semimajor axis of the inner planet as the scaling parameters. This is a matter of choice but helps maintain a certain homogeneity, regardless of the mass ratio under consideration. 

Fig.~\ref{fig6} shows the detectability limit combining results for all mass ratios. For a given $\sigma^*$, all detectable planetary systems lie inside a pair of corresponding curves. Note that the intersection between both sets of curves occurs for $m_2/m_1 \sim \sqrt[3]{2}$, as expected from the condition $K_2=K_1$. Finally, there is a certain variation with the orbital eccentricities stemming from the relation between signal amplitude and planetary mass (see equation (\ref{eq13bis})). However, qualitatively the picture remains the same. Except for very small standard deviations, it appears very difficult to detect resonant planetary systems with mass ratios $m_2/m_1$ larger than $\simeq 4$ or lower than $\simeq 0.3$. It is important to stress that this does not mean that it is impossible to detect systems outside these borders; but this detectability will not be guaranteed for any data set and will depend on the specific time distribution and/or errors of individual data points. 

The previous analysis assumes that $m_1 = M_{\rm Jup}$, $a_1 = 1$ AU and $m_0 = M_\odot$, for which $\sigma^* = \sigma$. Other values however, will change the picture. Among the known resonant population, perhaps the most extreme case is the GJ876 system ($m_2/m_1 \sim 3.1$), where $m_1 = 0.6 M_{\rm Jup}$, $a_1 = 0.13$ AU and $m_0 = 0.32 M_\odot$ (Butler et al. 2006). Applying equation (\ref{eq20cuatris}) to the results shown in Fig.~\ref{fig6}, we find that for $N=100$ the resonant system can be identified even for standard deviations in the RV data of the order of $\sigma \sim 40$ m s$^{-1}$ (corresponding to $\sigma^* \simeq 8$ m s$^{-1}$). 

\begin{figure}[t]
\begin{center}
\includegraphics*[width=18pc]{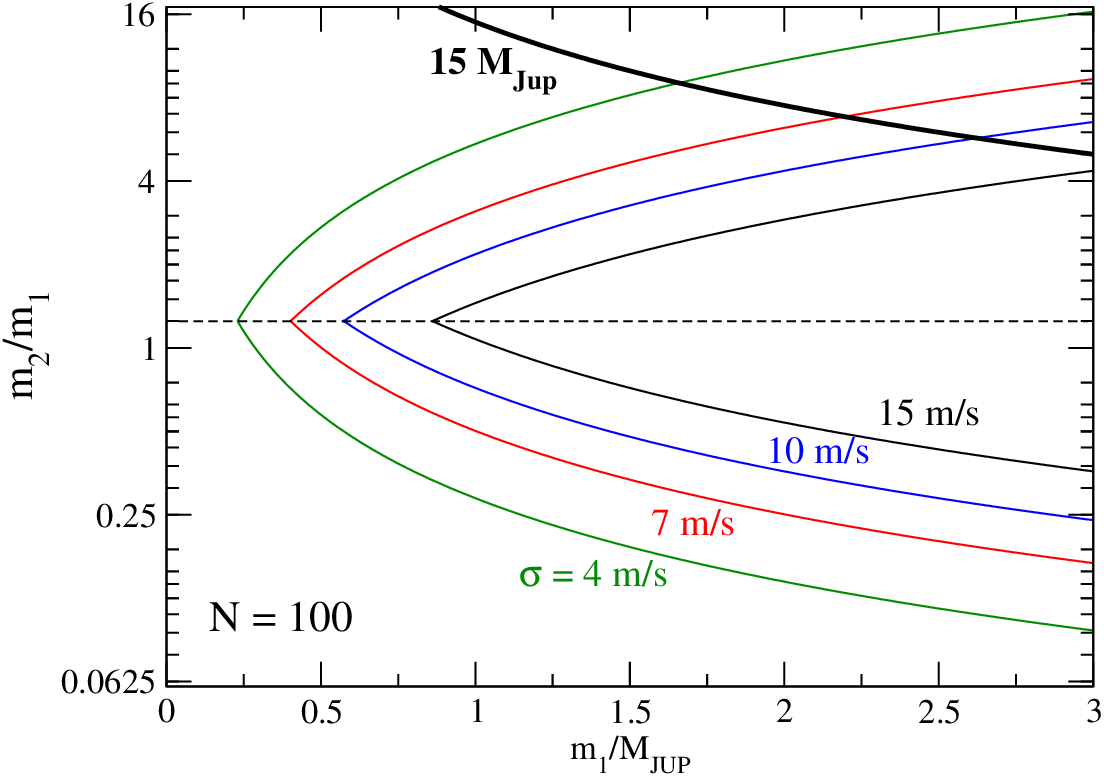}
\end{center}
\caption{Detectability limits of the mass ratio $m_2/m_1$ (with $P_{detect}= 0.99$ and $N=100$), as a function of the inner mass $m_1$, for four values of the standard deviation $\sigma$. Both planetary eccentricities are chosen as $e_i = 0.1$. The FAP is taken equal to ${\cal F} = 10^{-4}$. As before, for a given value of $\sigma$, detectability is guaranteed only inside the two corresponding curves. Above the broad black curve $m_2 > 15 M_{\rm Jup}$, a value usually associated with the brown dwarf limit.}
\label{fig6bis}
\end{figure}

Fig.~\ref{fig6bis} shows how the detectability limit of $m_2/m_1$ varies with the magnitude of $m_1$, now assuming a fixed length of the data set $N=100$. Both $a_1$ and $m_0$ remain at their original values. Once again, a resonant system is detectable if the mass ratio $m_2/m_1$ is located inside two curves of equal color. We can see that the detectability increases significantly with the $m_1$, allowing a larger range of mass ratios than previously shown. However, the possibility of very massive resonant planets must be considered with care since the augmented mutual perturbations may compromise the dynamical stability of the system.

The difficulty in detecting resonant planets with significantly different masses is in accordance with currently known systems. Recall that Fig.~\ref{fig1} shows the distribution of mass ratios against orbital period ratio for all consecutive planetary pairs in multi-planetary systems. All planetary systems near the 2/1 MMR have mass ratios $m_2/m_1$ grouped near unity, in accordance to the analytical findings of the present model. The interesting conclusion is that it is possible that resonant systems may exist for other mass ratios, but are presently undetectable.

\begin{figure}[t]
\begin{center}
\includegraphics*[width=18pc]{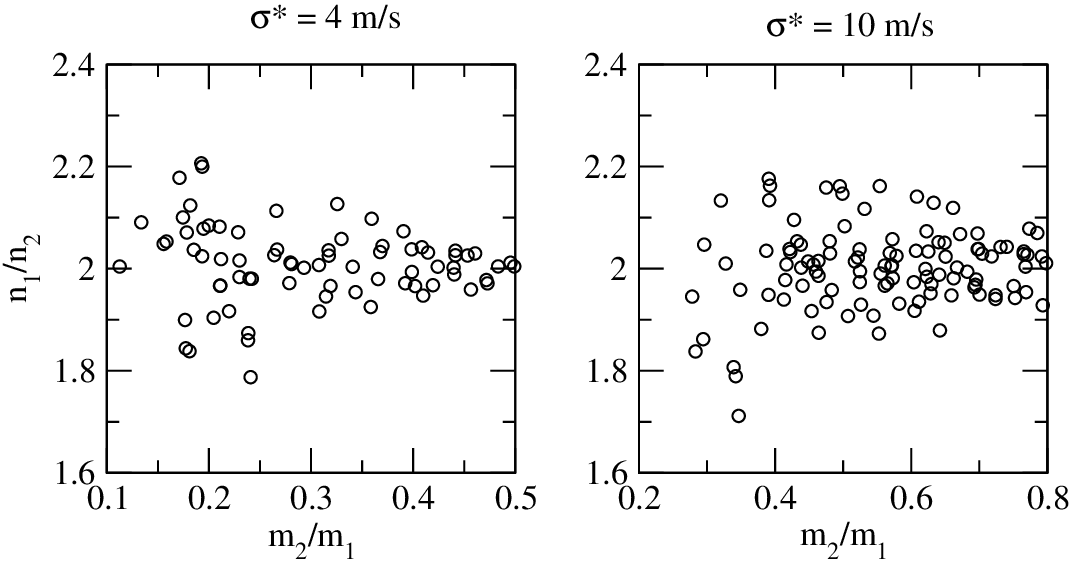}
\end{center}
\caption{Distribution of ratio of mean motions as function of planetary mass ratio $m_2/m_1$, from the Monte Carlo simulations used in Fig.~\ref{fig5}. Only those systems with ${\cal F} < 10^{-4}$ were used.}
\label{fig7}
\end{figure}

\section{Error Estimation in the Fitted Parameters}

Even if the two-planet resonant system is actually detected, there is no guarantee that the masses and orbital parameters can be estimated with any accuracy. As an example, Fig.~\ref{fig7} shows the distribution of ratios of mean motion for those synthetic data sets in Fig.~\ref{fig5} that satisfied the detectability criterion. A significant dispersion in the mean motions is noted around the nominal value $n_1/n_2=2$, with an appreciable percentage of the systems which would be qualified as near-resonant but not locked in MMR. This effect presents an additional difficulty in detecting resonant planets for mass ratios $m_2/m_1$ different than unity.

These results were obtained with synthetic data sets covering four orbital periods of the outer planet. Although the detectability criterion is independent of the total observational time interval (as long as we can guarantee a good coverage of the phases), the precision of the orbital elements is expected to improve with longer time intervals. Nevertheless, even if correctly identified as being locked in resonance, the planetary orbital elements may still contain significant errors and affect the deduced dynamics for the fitted systems. To quantify this effect, will analyze three test cases, one with mass ratio larger than one, one with equal masses and one with mass ratio lower than unity. 

\begin{figure}[t]
\begin{center}
\includegraphics*[width=18pc]{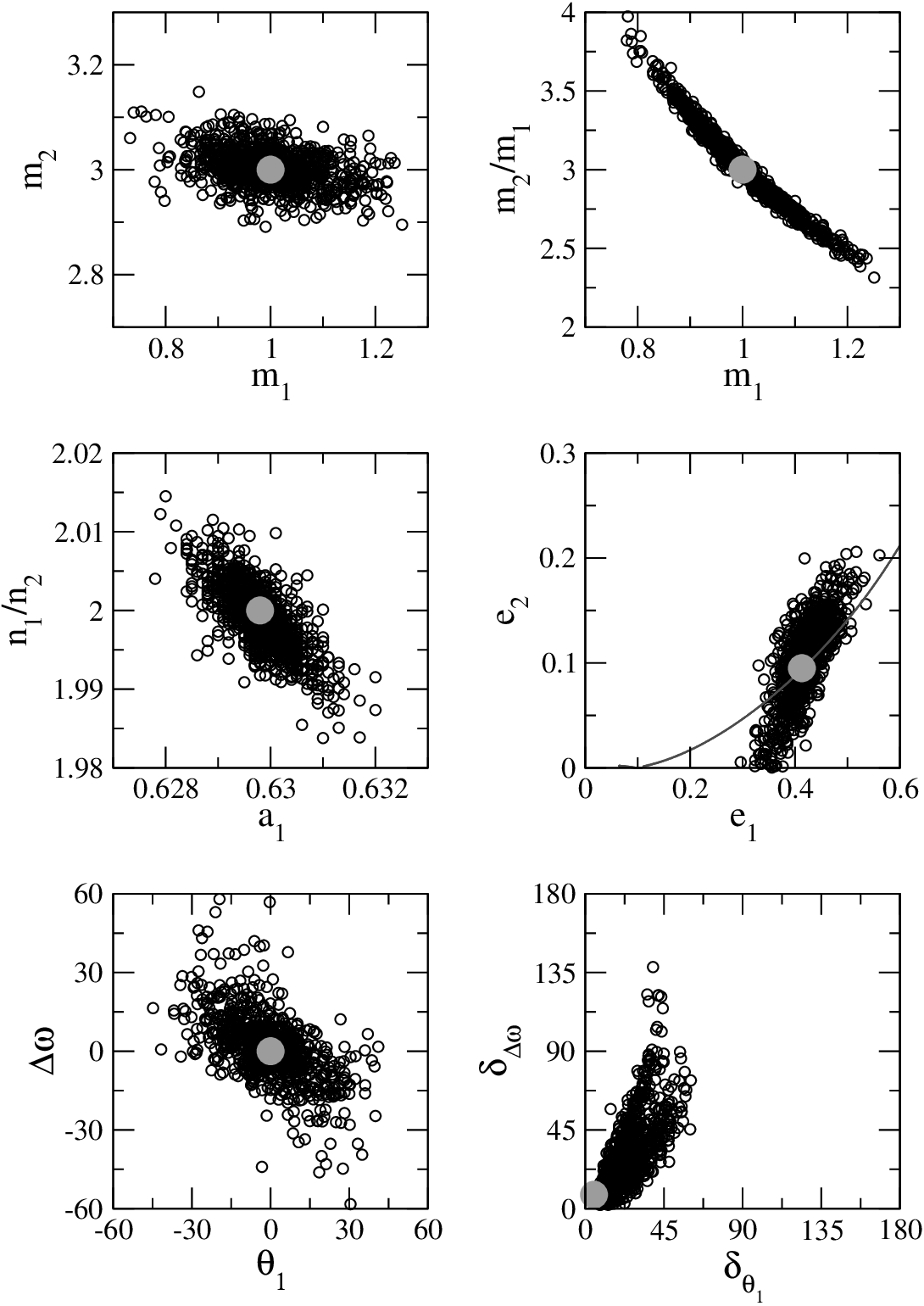}
\end{center}
\caption{Results of best fits of 1000 synthetic data sets corresponding to a nominal solution given by the first data set in Table \ref{table1}, with $N=200$ data points and constant standard deviation of $\sigma = 7$ m s$^{-1}$. The large filled gray circles indicate nominal solution. In the eccentricities plane the gray curve marks the family of zero-amplitude ACR solutions for a mass ratio $m_2/m_1=3/1$.}
\label{fig8}
\end{figure}

\subsection{More Massive Outer Planet}

We begin analyzing the case where the outer planet is more massive than its inner companion (i.e., $m_2/m_1>1$). We have chosen the first set of initial conditions shown in Table \ref{table1}, characterized by a mass ratio of $m_2/m_1=3/1$ and orbital elements leading to a small amplitude ($\sim 5^\circ$) oscillation around a $(0,0)$-type ACR. We generated 1000 synthetic RV data sets, each with $N=200$ data points distributed randomly over four orbital periods of the outer planet (i.e., four years). The values of each individual RV were taken randomly following a Gaussian distribution around the exact value and a variance given by a constant standard deviation for $\sigma = 7$ m s$^{-1}$ (i.e., $\sigma^* = 5.5$ m s$^{-1}$). These values guarantee a detectability of the two planets and places the deduced mean-motion ratio within the commensurability region. 

For each data set we first calculated a single-planet orbital fit starting with a genetic algorithm (Charbonneau 1995) without any pre-established initial solution. If the resulting fit had an associated FAP below $10^{-4}$ it was accepted and a two-planet fit was performed. The single-planet solution was used as a first guess and the code searched for the two-planet solution with the minimum rms. Once again this was accepted only if the FAP of the two-planet fit with respect to the single planet solution was below $10^{-4}$. If accepted, we then numerically integrated the configuration over a time span of $10^4$ orbital periods and estimated the amplitude of oscillation of the resonant angles $\theta_i = 2\lambda_2 - \lambda_1 - \varpi_i$ and the difference in longitudes of pericenter $\Delta \varpi = \varpi_2 - \varpi_1$. The critical angle $\theta_1$ corresponds to an interior resonance (see Michtchenko et al. 2008a), as found in symmetric ACR, while $\theta_2$ is the librating angle for exterior resonances (see Michtchenko et al. 2008b) and is thus applicable to asymmetric ACR.

Results are shown in Fig.~\ref{fig8}, where we present the distribution in several planes. In each frame the large filled gray circle marks the nominal solution and the open black circles the best fits of each synthetic data set. The gray curve in the eccentricities plane shows the $(0,0)$-family of ACR for a mass ratio $m_2/m_1$ equal to $3/1$. As expected from the choice of $N$ and $\sigma$, all fictitious two-planet systems were effectively detected by the fitting process and also identified as being locked in the 2/1 MMR, and both $\theta_1$ and $\Delta \varpi$ oscillated around the equilibrium solution. However, on $2\%$ of the cases $\Delta \varpi$ circulated taking all values between $0$ and $2 \pi$, while $\theta_1$ librated. This type of motion is usually referred to as a $\theta_1$-libration and is considered different from an ACR. However, as shown recently by Michtchenko et al. (2008a), even within an ACR the libration of $\Delta \varpi$ is purely kinematical and not associated with any separatrix crossing as long as the eccentricities are not too high.

The two top graphs in Fig.~\ref{fig8} show the distribution of planetary masses. From the left-hand plot we can see that the larger mass (i.e., larger $K$) has a fairly small dispersion of values, although $m_1$ is not very well determined leading to a large dispersion in the calculated mass ratio $m_2/m_1$ (right-hand graph) between $2.5$ and $3.5$. Even so, the semimajor axis and ratio of mean motion are very well estimated with little variation around the exact values (mid left-hand plot).

\begin{figure}[t]
\begin{center}
\includegraphics*[width=18pc]{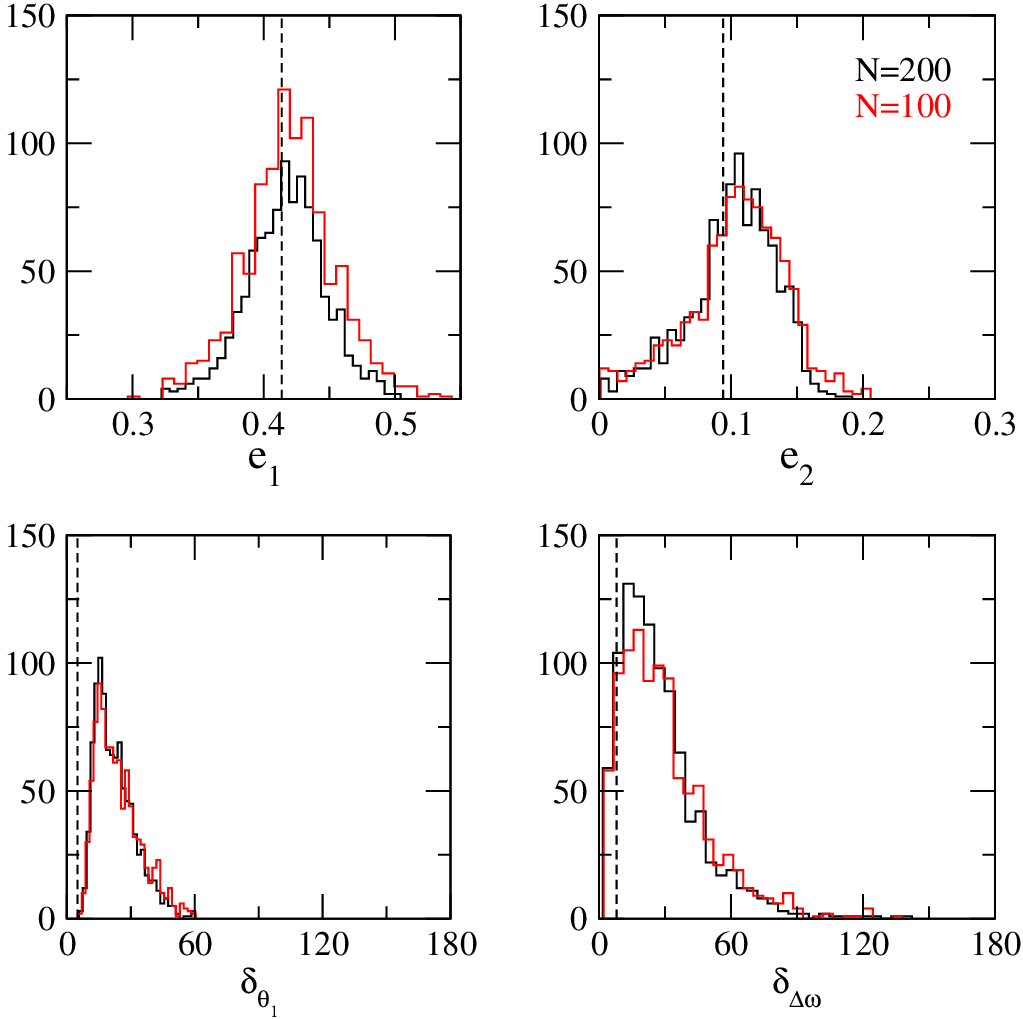}
\end{center}
\caption{Distribution of calculated eccentricities and amplitudes of oscillation of the resonant angle $\theta_1$ and $\Delta \varpi$ for the case $m_2/m_1=3/1$. Nominal values are indicated by the vertical dashed lines. The color codes are used to identify different sizes $N$ of the data sets.}
\label{fig9}
\end{figure}

The distribution of solutions in the eccentricities plane shows an interesting correlation with the family of zero amplitude ACR (shown in gray). This is an unexpected result since there is no evident relationship between the fitting process and the topology of the MMR. However, it does indicate that even imprecise determination of the eccentricities will not necessarily lead to large displacement from the ACR family and, consequently, from stable solutions. Recall that the successive orbital determinations of the resonant GJ876 planets (mass ratio close to $m_2/m_1=3/1$) show a similar behavior. Although in the past eight years the eccentricities have varied significantly, all best fit solutions have been found close to a stable ACR. 

Finally, the two bottom frames show the variation in the angular variables. The graph on the left gives the distribution of the calculated values of $\theta_1$ and $\Delta \varpi$. More interestingly, the right-hand frame shows the distribution of the amplitudes of oscillation around the ACR. Note a very important dispersion in the dynamical behavior of the calculated fits. As with the eccentricity, the amplitude has a lower bound at zero which causes a systematic effect leading to the fact that the estimated amplitude of libration is always overestimated with respect to the true value. 

This figure shows that, even under favorable circumstances given by a large data set and reasonable observational uncertainties, the errors in the orbital fit are still significant, especially in the eccentricities and angular variables. Even if the ACR is still identified as such, the amplitudes of oscillation are greatly increased beyond their nominal values. 

Fig.~\ref{fig9} shows the distribution of fitted eccentricities and amplitudes of oscillation of the angular variables for $N=200$ (black histograms) and $N=100$ (red histograms). The nominal values corresponding to the original system are marked with the vertical dashed lines. In all cases the data points were distributed randomly over four orbital periods of the outer planet and $\sigma=7$ m s$^{-1}$. 

\begin{figure}[t]
\begin{center}
\includegraphics*[width=18pc]{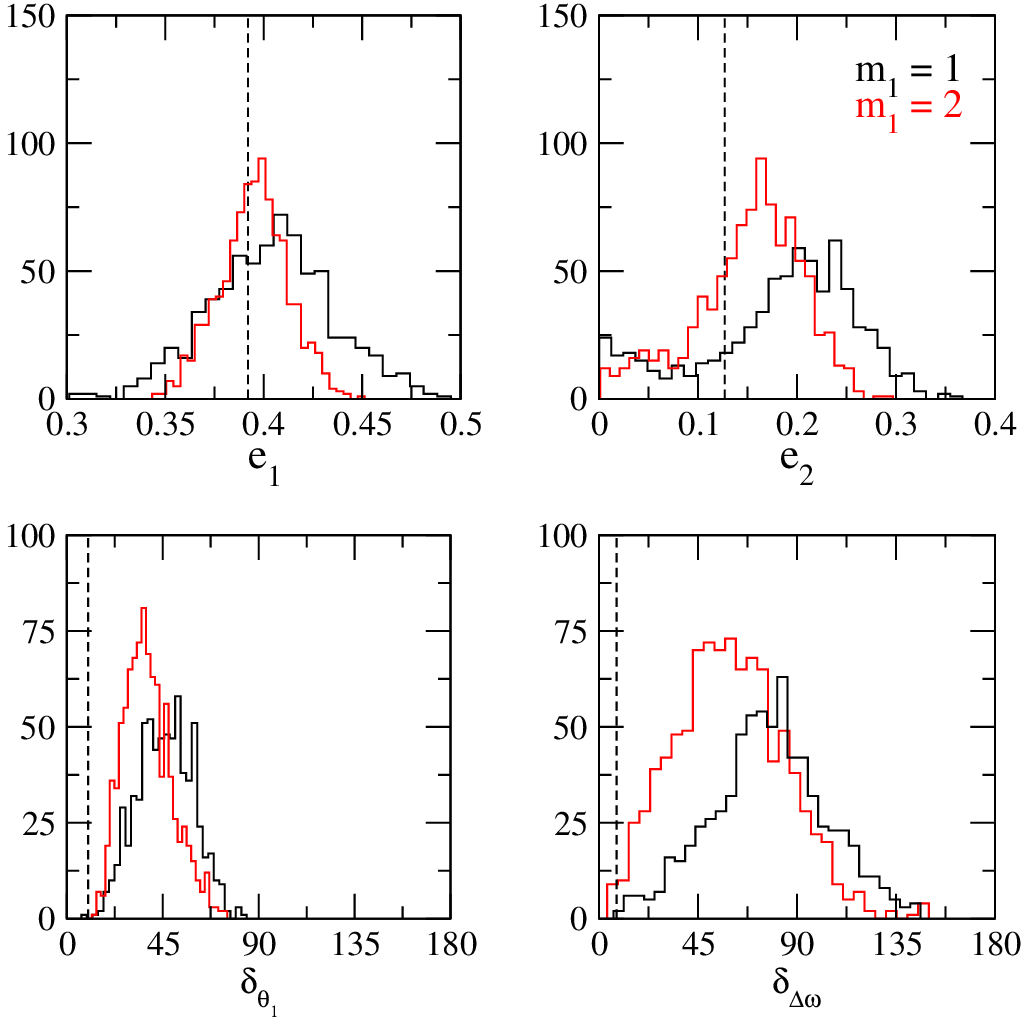}
\end{center}
\caption{Same as Fig.~\ref{fig9} for mass ratio $m_2/m_1 = 1/1$. However, now color codes are used to identify different values of $m_1$, maintaining a constant length of the data set $N=100$. The black histograms were obtained with $m_1 = 1 M_{\rm Jup}$ while those in red were calculated with $m_1 = 2 M_{\rm Jup}$.}
\label{fig10}
\end{figure}

The two top frames correspond to the eccentricities; $e_1$ on the left and $e_2$ on the right. The distribution of eccentricities of the inner planet is fairly symmetric with respect to the true value ($\sim 0.42$), and is reminiscent of the results shown for single planet systems (Shen \& Turner 2008). Although the dispersion increases slightly for smaller $N$, there appears to be little systematic error and the peak of the histogram is close to the vertical dashed line. The same, however, is less evident in the case of the eccentricity of the outer planet, where the distribution is less symmetric and there is a bias towards larger values of $e_2$.

The two lower plots give the distribution of amplitudes of oscillation of the resonant angle $\theta_1$ (left-hand graph) and $\Delta \varpi$ (right-hand graph). Once again the nominal values are shown in the vertical dashed lines and correspond to a small amplitude solution. As shown in Fig.~\ref{fig8}, there is an important increase in the amplitudes leading to a systematic bias towards solutions with larger amplitudes of oscillation. Surprisingly there is little change for smaller data sets. In the case of $\theta_1$ there is practically no calculated solution with amplitude larger than $90^\circ$. The same, however, does not hold for $\Delta \varpi$. Even for larger data sets, some solutions have amplitude equal to $180$ degrees corresponding to circulations of the difference in longitudes of pericenter.

\subsection{Equal Mass Planets}

The previous analysis can be repeated for a mass ratio of unity ($m_2/m_1=1$) and choosing initial conditions close to a stable ACR. Orbital elements are once again indicated in Table \ref{table1}. This particular ACR was chosen such that the total angular momentum of the system was approximately the same as in the previous case; mainly this implies that the sum of the square of the eccentricities is of the same magnitude. 

Fig.~\ref{fig10} shows histograms with the distribution of eccentricities (top), amplitude of libration of $\theta_1$ (bottom left) and amplitude of oscillation of $\Delta \varpi$ (bottom right). Contrary to Fig.~\ref{fig9}, the number of data points has been kept constant at $N=100$ and the color code now identifies different values for the mass of the inner planet. The black histograms were drawn using $m_1=M_{\rm Jup}$, while for the plots in red we adopted $m_1=2 M_{\rm Jup}$. This latter value generates a RV amplitude in the complete signal similar to the case discussed in Section 4.1, thus allowing an easier comparison of the dispersion of the orbital solutions.

As before, the eccentricity of the inner planet is fairly well established. For both values of $m_1$ the distribution is centered near the nominal value, although we note a smaller dispersion in the results for larger masses (i.e., larger amplitudes of the RV signal). The eccentricity of the outer planet, however, shows a more complex distribution, with a notorious systematic shift in the averaged value. Also as before, practically all the orbital fits lead to large amplitudes of libration, especially of the difference of longitudes of pericenter. The mean value of this last angle lies near $90^\circ$. Although these bias decrease significantly with larger values of $m_1$, they are still much larger than those presented in Fig.~\ref{fig9}.

\begin{figure}[t]
\begin{center}
\includegraphics*[width=18pc]{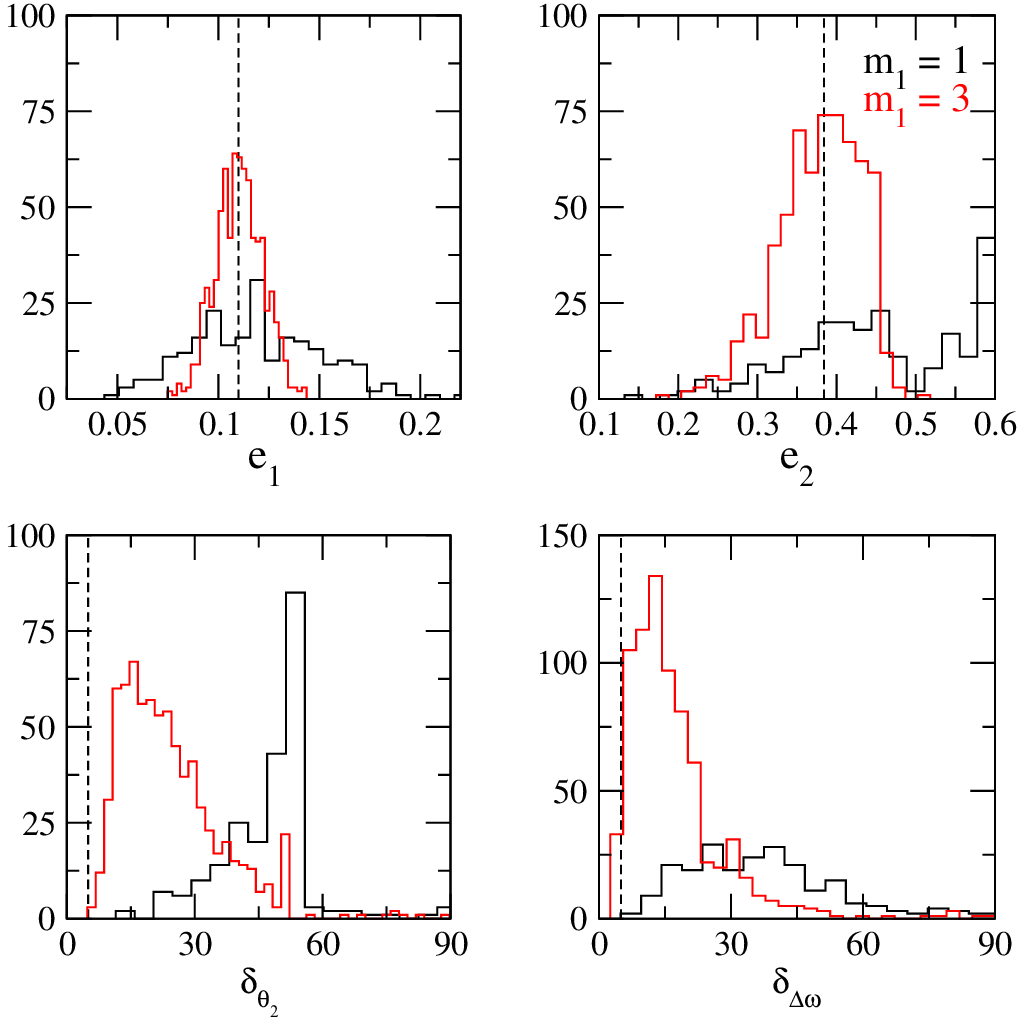}
\end{center}
\caption{Same as Fig.~\ref{fig10} for mass ratio $m_2/m_1 = 1/3$. Black histograms were obtained with $m_1 = 1 M_{\rm Jup}$ while those in red with $m_1 = 3 M_{\rm Jup}$.}
\label{fig11}
\end{figure}

\begin{table}[t]
\begin{center}
\small\addtolength{\tabcolsep}{-5pt}
\begin{tabular}{|c|l|c|c|c|}
\hline 
$m_2/m_1$   &            & Non resonant ($\%$) & $\theta_i$-Librator ($\%$) &  ACR ($\%$) \\
\hline 
   3/1      & $N = 200$    &   $0 $    &      $2 $         &  $98 $  \\
            & $N = 100$    &   $0 $    &     $12 $         &  $88 $  \\
\hline 
   1/1      & $m_1 = 1$  &   $9 $    &     $21 $         &  $70 $  \\
            & $m_1 = 2$  &   $2 $    &      $2 $         &  $96 $  \\
\hline 
   1/3      & $m_1 = 1$  &   $68 $    &     $6 $         &  $26 $  \\
            & $m_1 = 3$  &   $23 $    &     $1 $         &  $76 $  \\
\hline 
\end{tabular} 
\end{center}
\caption{Percentage of Different Dynamical Outcomes of the Best Fits for Different Mass Ratios $m_2/m_1$. For symmetric ACR the resonant angle is taken as $\theta_1$, while $\theta_2$ is adopted for asymmetric solutions (see Michtchenko et al. 2008ab). The values of $m_1$ are given in units of $M_{\rm Jup}$.}
\label{table2}
\end{table}

\subsection{Less Massive Outer Planet}

Finally, we analyze the case where $m_2/m_1<1$. Initial conditions (Table \ref{table1}) correspond to a small amplitude oscillation around an asymmetric ACR with $m_2/m_1=1/3$. Once again the histograms in red were obtained with increased planetary masses ($m_1 = 3 M_{\rm Jup}$) which yield a magnitude in the complete RV signal similar to the initial conditions for $m_2/m_1=3/1$. 

Although both planets were detected in all the data sets, for $m_1 = M_{\rm Jup}$
only $26 \% $ of the resulting orbits reproduced the ACR. In $68 \%$ of the cases the resulting configuration was non-resonant and neither $\theta_2$ nor $\Delta \varpi$ librated around the equilibrium values given by the ACR, but instead circulated taking all values between zero and $2 \pi$. Given the high eccentricity of the outer planet, these orbits inevitably led to unstable motion and a disruption of the system in short timescales. In the remaining $6 \%$ of the cases 
the best fits yielded $\theta_2$-librations but a circulation of the difference in longitudes of pericenter.

These numbers improve significantly for $m_1 = 3 M_{\rm Jup}$, where disruptions occurred only in $23 \% $ of the cases and most of the rest corresponded to stable ACR. Even so, this implies that for scale standard deviations as low as $\sigma^* = 2$ m s$^{-1}$, a quarter of the synthetic data sets failed to identify the resonant motion and led to dynamically unstable fits. 

The distribution of the orbital elements of the stable ACR solutions is shown in Figure \ref{fig11}. Since we have eliminated the non resonant cases, the histograms appear similar to the other mass ratios, with a fair estimation of the eccentricity of the inner planet and a much larger dispersion in $e_2$. For $m_1= M_{\rm Jup}$ we also note indications of a bimodal distribution for the eccentricity of the outer planet, with one peak centered approximately around $e_2 = 0.45$ and a secondary peak near $0.6$. This bi-modality is related to the bifurcation of the family of asymmetric ACR for this mass ratio and high eccentricities (see Michtchenko et al. 2008b). However, the histogram in red shows a more localized distribution of $e_2$.

\begin{figure}[t]
\begin{center}
\includegraphics*[width=13pc]{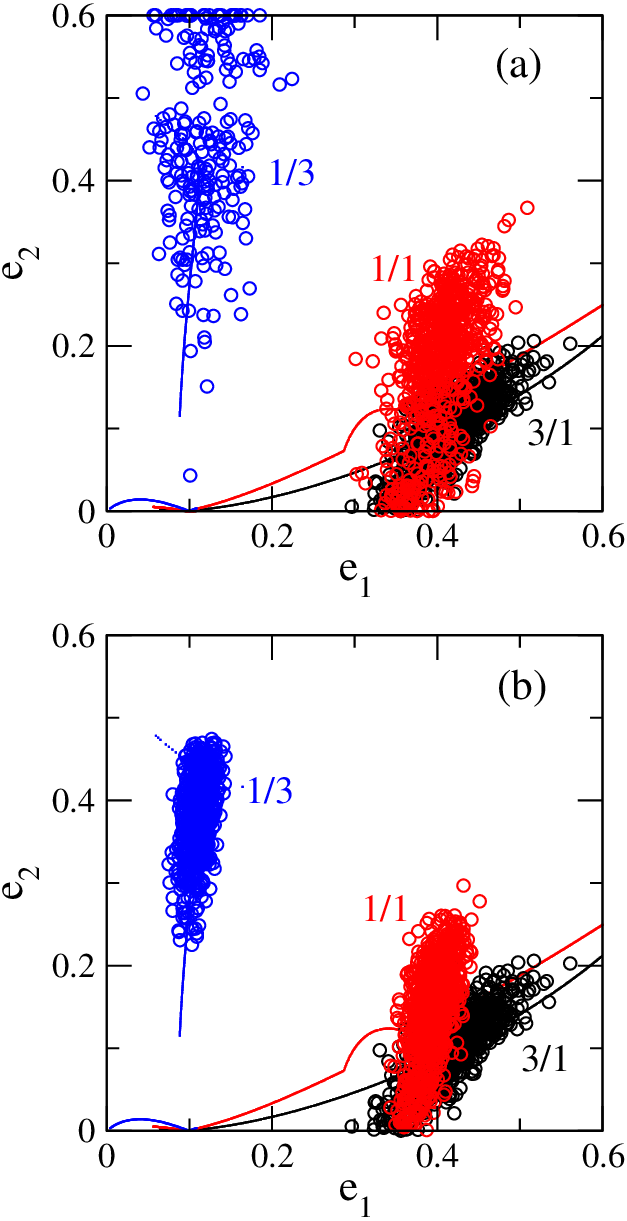}
\end{center}
\caption{Distribution of planetary eccentricities from best fits of synthetic data sets, compared with the families of ACR. The color codes identify different mass ratios. Black : $m_2/m_1 = 3/1$, red : $m_2/m_1 = 1/1$ and blue : $m_2/m_1 = 1/3$.
{\bf (a)}: All inner planetary masses equal to $1 M_{\rm Jup}$. {\bf (b)}: Augmented inner planetary masses: $m_1 = 2 M_{\rm Jup}$ for $m_2/m_1 = 1/1$ and $m_1 = 3 M_{\rm Jup}$ for $m_2/m_1 = 1/3$.}
\label{fig12}
\end{figure}
 
As with the other mass ratios, the distributions of amplitudes of libration/oscillation also show a marked bias towards larger values. Although this decreases for larger values of $m_1$, even for $m_1=3M_{\rm Jup}$ practically none of the best fits are able to reproduced the small-amplitude oscillation around the asymmetric ACR. Finally, in the black histogram the distribution of the amplitude of libration of $\theta_2$ shows a marked peak around $60$ degrees, which roughly corresponds to the maximum amplitude permitted by the asymmetric ACR. Larger amplitudes place the system outside the resonant domain, which may explain the large proportion of non-resonant outcomes in the orbital fits.

\subsection{Application to Real Exosystems}

Table \ref{table2} summarizes the possible dynamical outcomes of the orbital fits for the three mass ratios. Although all data sets were constructed with the same standard deviation $\sigma=7$ m s$^{-1}$, there is a marked difference in the results. For $m_2/m_1=3/1$ most fits correctly identified the ACR and all led to stable motion of both planets in timescales of the order of $10^4$ orbital periods. The inverse case, $m_2/m_1=1/3$ is significantly less precise, and is probably due to the existence of several domains of stable motion, each associated with a different family of ACR (Michtchenko et al. 2008b). Each domain is separated by a chaotic layer, and are less robust than those found for mass ratios larger than unity. 

It thus appears that resonant planets with mass ratios lower than unity are not only more difficult to detect, but even when detected the orbital fits usually lead to wrong period ratios and/or dynamically unstable solutions. Although these results are valid for the 2/1 resonance and the 55Cnc-c and 55Cnc-d planets are located in the vicinity of the 3/1 commensurability (mass ratio $m_2/m_1 \simeq 1/5$), it seems plausible to expect similar results, particularly since asymmetric ACR are involved in both cases. This could be a possible explanation of why successive orbital fits of the 55Cnc planets alternatively point towards resonant or non-resonant motion, while the same uncertainty is not present for exosystems with other mass ratios. Another example is given by the HD37124 system with two planets in the vicinity of the 2/1 MMR. Although the best fit indicates $m_2/m_1 \simeq 1.2$ and an asymmetric ACR (Baluev 2008), other solutions exist in which the bodies are outside this commensurability. 

Fig.~\ref{fig12} compares the distribution of eccentricities found for $m_2/m_1=3/1$ (black open circles), $m_2/m_1=1/1$ (red) and $m_2/m_1=1/3$ (blue). Only those orbital fits leading to stable ACR were plotted. The colored curves correspond to the zero-amplitude ACR for each mass ratio. The top plot was drawn using $m_1 = M_{\rm Jup}$ in all three cases. For the bottom plot we used the augmented values of $m_1$ given in the caption. 

For $m_2/m_1=1/3$ we note that most of the solutions lie relatively close to the family of stable ACR, although it is important to recall that these correspond to less than half of the total data sets. A similar trend is observed for $m_2/m_1=3/1$, except that all orbital fits were now plotted. However, in the case of equal masses $m_2/m_1=1/1$, there is no apparent correlation between the eccentricity dispersion and the ACR families is absent in this case. Thus, it appears that, for large mass ratios, there is a much larger probability that even uncertain orbital fits will remain close to stable ACR configurations. For mass ratios close to unity, this is in no way guaranteed, and even small standard deviations may lead to larger amplitude of libration or unstable motion. Even with increases planetary masses and, thus, better orbital fits, the same correlation is still found. In fact, the large signal-to-noise ratio of the RV data maintains a similar dispersion in $e_2$, although the estimated value of $e_1$ is more precise.

\section{Conclusions}

We have analyzed the detectability limit of two planets in the vicinity of a 2/1 MMR, as a function of the number $N$ of data points and the standard deviation $\sigma$ of the RV values. For values of $\sigma$ and $N$ comparable to real exosystems we found an important bias toward cases with $m_2/m_1 \sim 1$. Planets with mass ratios much larger or smaller than unity are much more difficult to separate from Doppler data. This may not only help explain the lack of resonant systems with these mass ratios but also may indicate that the number of resonant systems in existence is actually significantly larger than currently believed.

We also analyzed the dispersion in the masses and orbital elements of the detected systems. Although for large mass ratios most synthetic data sets effectively detect a resonant motion, for low values of $m_2/m_1$ a large proportion of the orbital fits fail to detect the resonance lock and erroneously indicate a non-resonant configuration. Even within the subset of resonant orbits, we note a large dispersion in the estimated values of the eccentricity of the outer planet, and the median of the distribution is always higher than the nominal value of $e_2$. This is in accordance with several real exosystems in the 2/1 MMR, particularly HD82943. Moreover, nominal solutions corresponding to small-amplitude librations around ACR are usually accompanied by an important bias toward incorrect large amplitudes that may even compromise their dynamical stability on long timescales. 

Even if this is not the case, it presents a possible answer to the question of why, with the exception of the resonant bodies in GJ876, all other resonant planets have orbital fits consistent with large amplitude librations. Although recent works by S\'andor, et al (2007) and Crida et al. (2008) propose that the observed large amplitudes of libration may have been caused by past dynamical processes (e.g., planetary scattering or disk dispersal), the present results indicate that this observational characteristic may not be real, but a simple consequence of dealing with small data sets with large uncertainties in the RV values. 

For multiplanetary systems it is important to stress that the absolute numerical precision of the orbital elements is not, in itself, the only datum to consider, as much as the diversity in dynamical behaviors within the uncertainty region. In other words, if the two planets are non resonant with low eccentricities, even significant changes in the orbital elements will not necessarily lead to different regimes of motion. Conversely, if the planets are locked in a MMR and (preferably) have moderate to high eccentricities, even a small change in the orbital parameters may mean the difference between stable and unstable motion.

\acknowledgements
{\bf Acknowledgments:} This work has been supported by the Argentinean Research Council - CONICET, the Brazilian National Research Council - CNPq, and the S\~ao Paulo State Science Foundation -FAPESP-. The authors also gratefully acknowledge the support of the CAPES/Secyt program for scientific collaboration between Argentina and Brazil.

\section*{Appendix. Data fitting to a model with orthonormal functions.}

We assume that the adopted model ${W}(t)$ is given by a linear combination of $M$ independent continuous functions $H_\mu(t)$ ($\mu=0,1,\cdots,M-1$). This is, for instance, the case of the very simple model of one planet in circular orbit, for which the true and mean anomalies are equal $f=\ell$ (for a given fixed period). Although the general (elliptic orbit) model is no longer linear, we can linearize it in the neighborhood of the true values, which are then assumed to be determined by successive iterations. We assume that the functions $H_\mu(t)$ can be orthonormalized to generate a new set of $M$ functions $h_\mu(t)$ such that 
\be
(h_\mu,h_\rho)=\delta_{\mu,\rho}
\ee
where $\delta_{\mu,\rho}$ is the Kronecker symbol and $(h_\mu,h_\rho)$ represents the inner product 
\be
(h_\mu,h_\rho) = \sum_1^N h_\mu(t_i) h_\rho(t_i).
\ee
If the observations do not have the same variance, weights may be introduced in this definition to take it into account. 
Hence
\be
{W}(t_i)=\sum_0^{M-1} {\beta}_\mu H(t_i)
=\sum_0^{M-1} {\alpha}_\mu h(t_i)
\ee
(see Kurth, 1967). The construction of the orthonormal functions is a cumbersome step. In the case of the search of periods of unequally spaced observations, they were first introduced by Ferraz-Mello (1981). In this application, it is not necessary to actually construct the orthonormal functions $h(t_i)$. It is only necessary to know that such operation is possible and that these functions allow much easier derivations as they behave just as orthogonal unit vectors. In the same way, we do not need here the actual definition of the $\alpha$'s (as functions of the $\beta$'s).

One important point often overlooked is that the set of function $H_\mu(t)$, for reasons of completeness, must include the function $H_0(t)=1$. In  Scargle's (1982) periodogram, for instance, this function is not included.  However, the assumption that the given data have zero average (i.e., $\sum_1^N y(t_i) = 0$) does not guarantee that the data modeled at the given times will also have zero average. In general, if ${W}(t_i)$ are the radial velocities computed at the times $t_i$ using the proposed model, we have $\sum_1^N {W}(t_i) \neq 0$. 
Consequently an incomplete basis, in which $H_0(t)=1$ is missing, cannot allow the function that model the data to be represented. When just the period of such a function is looked for (the usual question that a periodogram analysis is assumed to answer), a basis formed by $\sin(\omega t)$ and $\cos(\omega t)$ may be enough, but for the full problem of best fitting one sinusoid to the data, it is not. For this reason, in many applications, as those concerning ``detectability" using periodograms, the complete model given by Ferraz-Mello (1981) (or some equivalent ones published later; see references in Cumming et al. 1999) must be used instead of the incomplete ones. 

The following results are elementary:

\begin{itemize}

\item The solution $\widehat{Q}=Q_{\rm min}$ of the best fit problem is given by
\be
\widehat{\alpha}_\mu=(V,h_\mu) ,
\ee
which is obtained from $\partial Q/\partial \alpha_\mu=0$. Note that $h_0=1/N$ and, therefore, $\widehat{\alpha}_0=\langle V \rangle$ ;

\item In the best fit, the square sum of the residuals is given by 
\be
\widehat{Q} = ||V||- ||\widehat{W}||
\label{eqq}
\ee
where $||V|| \equiv (V,V)$ and $||\widehat{W}||$ is the spectral power 
\be
||\widehat{W}|| \equiv (\widehat{W},\widehat{W}) = \sum_0^{M-1} \widehat{\alpha}_\mu^2 ;
\ee

\item If we assume that the data are $V_i = {\mathcal{W}}(t_i) + \epsilon_i$ where $\mathcal{W}$ is a given model function and $\epsilon_i$ a normal variate $N(0,\sigma^2_\epsilon)$ uncorrelated with $\mathcal{W}$ (i.e., $(\mathcal{W},\epsilon)=0$), there follows $||V||=||{\mathcal{W}}|| + ||\epsilon||$ and, therefore, 
\be
||\widehat{W}|| = ||{\mathcal{W}}||+||\epsilon||-\widehat{Q}\,;
\label{normadeW}
\ee

\item If the observations are perfectly distributed,
\be
||\mathcal{W}|| = \sum \mathcal{W}_i^2 = N\langle \mathcal{W}^2 \rangle =\lim_{T\rightarrow\infty} \frac{N}{T} \int_0^T \mathcal{W}(t)^2dt.
\ee
In the case of one planet in circular orbit, this operation gives the value $NK^2/2$ found in the quoted papers. However, even in this more simple case, if the observations are not well distributed, the non-orthonormality of the functions $H_\mu(t)$ leads to a different result.

\end{itemize}

Note that when the given data averages to zero: $\langle V \rangle = 0$. This often occurs because it is usual in periodogram studies to refer the given data to its average. In such case $\widehat{\alpha}_0=0$ and $||V||=(V,V)$ is the variance of the given data ($V_i$) times $(N-1)$.

Two particular cases are important in the application to exoplanetary detection.

\begin{enumerate}
\item 
One-Planet Case:

Here model $a$ is a constant (i.e., ${W_a}={\beta}_0$). Then we can simply write $||\widehat{W}_a||=\alpha_0^2=\langle V \rangle^2$. If, in addition, 
the given data are such that $\langle V \rangle = 0$, then the floating mean power ${z}$ (see equation (\ref{z-floating})) corresponding to the best fit solution is
\be
\widehat{z} = \frac{N-M_b}{M_b-1}\,\frac{||\widehat{W}_b||}
{\widehat{Q}_b}.
\label{zsimples}
\ee
Introducing (\ref{normadeW}) we obtain
\be
\widehat{z} = \frac{N-M_b}{M_b-1}\,
\biggl( \frac{||{\mathcal{W}_b}||+||\epsilon||}{\widehat{Q}_b} - 1\biggr).
\label{zduplo_antes}
\ee
Considering that $||\epsilon|| = \sum_i \epsilon_i^2$ is an estimation of 
$(N-1) \, \sigma^2_\epsilon$, where $\sigma_\epsilon$ is a characteristic standard deviation of the RV data, then finally
\be
\widehat{z} = \frac{N-M_b}{M_b-1}\,
\biggl( \frac{||{\mathcal{W}_b}||+(N-1) \sigma^2_\epsilon}{\widehat{Q}_b} - 1
\biggr),
\label{zduplo}
\ee
where recall that ${\mathcal{W}_b}$ is the signal generated by the single planet.

\item 
Two-Planet Case:

We now assume that model \textit{a} is the one-planet model and \textit{b} the two-planet model. As before, the synthetic sample is given by $V_i = \mathcal{W}_b(t_i)+\epsilon_i$ where $\epsilon_i$ is a Gaussian noise. We assume that $\mathcal{W}_b(t_i)$ is the sum of two one-planet models $\mathcal{W}_1(t_i)$ and $\mathcal{W}_2(t_i)$, where the index are chosen such that the signal of the first planet is larger (i.e., $||{\mathcal{W}_1}|| > ||{\mathcal{W}_2}||$). We can then write
\be 
||V|| = ||\mathcal{W}_1|| + ||\mathcal{W}_2|| + (\mathcal{W}_1,\mathcal{W}_2) + ||\epsilon||
\ee
where $(\mathcal{W}_1,\mathcal{W}_2)$ is the covariance between the signals generated by each individual planet. Introducing this expression into (\ref{eqq}),
we can write
\be 
\widehat{Q}_a = ||\mathcal{W}_1|| + ||\mathcal{W}_2|| + (\mathcal{W}_1,\mathcal{W}_2) + ||\epsilon|| - ||\widehat{W}_a||
\ee
which can be substituted into
\be
\widehat{z} = \frac{N-M_b}{M_b-M_a}\,\frac{\widehat{Q}_a-\widehat{Q}_b}
{\widehat{Q}_b}.
\label{zd}
\ee
to yield
\be 
\widehat{z} = \frac{N-M_b}{M_b-M_a}\, \biggl( \frac{||\mathcal{W}_2|| +(N-1)\sigma_\epsilon'^2}{\widehat{Q}_b}-1 \biggr)
\label{zfin}
\ee
where note that the expression depends solely on the signal generated by the planet with the weakest contribution. Its form is analogous to the one-planet case, where the modified variance $\sigma_\epsilon'^2$ is defined as
\be 
\sigma'^2_{\epsilon} = \sigma^2_{\epsilon} + \frac{(\mathcal{W}_1,\mathcal{W}_2)}{(N-1)} + \frac{(||\mathcal{W}_1||-||\widehat{W}_a||)}{(N-1)} .
 \ee
In principle, the two additional terms in the right-hand side cannot be set to zero. First, the contribution from the two planets to the signal cannot be separated and when $||\mathcal{W}_a||$ is obtained by means of an one-planet fit, the result is certainly contaminated by the fact that there is a second planet contributing to the signal $||V||$. Second, the covariance $(\mathcal{W}_1,\mathcal{W}_2)$ may be different from zero, especially if the two planets are assumed to be in resonance. However, both terms are inversely proportional to $(N-1)$, and it is expect that $\sigma'^2_{\epsilon} \rightarrow \sigma^2_{\epsilon}$ for sufficiently large data sets.
\end{enumerate}


\begin{thebibliography}{}

\bibitem[]{aea08}
Anglada-Escud\'e, G., L\'opez-Morales, M. \& Chambers, J.E., 2008. submitted to ApJL., arXiv:0809.1275.

\bibitem[]{ad08}
Adams, F.C., Laughlin, G. \& Bloch, A.M., 2008. ApJ, 683, 1117.

\bibitem[]{b08}
Baluev, R. V. 2008, Celest. Mech. Dyn. Astron., 102, 297.

\bibitem[]{bmf06}
Beaug\'e, C., Michtchenko, T.A. \& Ferraz-Mello, S., 2006. MNRAS, 365, 1160.

\bibitem[]{bfm07}
Beaug\'e, C., Ferraz-Mello, S. \& Michtchenko, T.A., 2007, In: {\it Extrasolar Planets: formation, Detection and Dynamics ed.R.Dvorak} (Weinheim: Wiley-V CH), 1-25 

\bibitem[]{bgfm08}
Beaug\'e, C., Giuppone, C.A., Ferraz-Mello, S. \& Michtchenko, T.A., 2008. MNRAS, 385, 2151.

\bibitem[]{but06}
Butler, R. P., et al. 2006. ApJ, 646, 505.

\bibitem[]{char95}
Charbonneau, P., 1995. ApJS, 101, 309.

\bibitem[]{cel09}
Correia, A.C.M., et al. 2009, A\&A, 496, 521

\bibitem[]{cea08}
Crida, A., S\'andor, Zs. \& Kley, W., 2008. A \& A, 483, 325.

\bibitem{cea99}
Cumming, A, Marcy, G.W. \&  Butler, R.P., 1999. ApJ. 526, 890.

\bibitem[]{c04}
Cumming, A., 2004. MNRAS, 354, 1165.

\bibitem[]{dal08}
D'Angelo, G. \& Lubow, S.H., 2008. ApJ, 685, 560.

\bibitem{sfm}
Ferraz-Mello, S., 1981. AJ., 86, 619.

\bibitem{FMQ}
Ferraz-Mello, S. \& Quast, G., 1987. In {\it Exercises in Astronomy} (J. Kleczek, ed.), D. Reidel, Dordrecht, 231.

\bibitem[]{fea08}
Fischer, D.A., et al. 2008, ApJ, 675, 790.

\bibitem[]{gel05}
Go\'zdziewski, K., Konacki, M. \& Maciejewski, A.J., 2005. ApJ, 622, 1136.

\bibitem[]{h08}
Hadjidemetriou, J.D. 2008, Celest. Mech. Dyn. Astron., 102, 69.

\bibitem[]{gel07}
Go\'zdziewski, K., Maciejewski, A.J. \& Migaszewski, C., 2007. ApJ, 657, 546.

\bibitem[]{lea06}
Kley, W., Lee, M.H., Murray, N. \& Peale, S.J., 2005. A \& A, 437, 727.

\bibitem{Kur}
Kurth, R., 1967. {\it Introduction to Stellar Statistics}, Pergamon Press, Oxford.

\bibitem[]{lc09}
Laskar, J. \& Correia, A.C.M., 2009. A\&A, 496, 5L

\bibitem[]{lp02}
Lee, M.H. \& Peale, S.J., 2002. ApJ, 567, 596.

\bibitem[]{sn}
Masset, F. \& Snellgrove, M. 2001, MNRAS, 320, L55.

\bibitem[]{mbf08a}
Michtchenko, T.A., Beaug\'e, C \& Ferraz-Mello, S., 2008a. MNRAS, 387, 747.

\bibitem[]{mbf08b}
Michtchenko, T.A., Beaug\'e, C \& Ferraz-Mello, S., 2008b. MNRAS, 391, 227.

\bibitem[]{mor}
Morbidelli, A. \& Crida, A., 2007. Icarus, 191, 158.

\bibitem[]{ma05}
Moorhead, A.V. \& Adams, F.C., 2005. Icarus, 178, 517.

\bibitem[]{sea06}
S\'andor, Zs., Kley, W. \& Klagyivik, P., 2007. A \& A, 472, 981.

\bibitem[]{s82}
Scargle, J.D., 1982. ApJ, 263, 835.

\bibitem[]{st08}
Shen, Y. \& Turner, E.L., 2008. ApJ, 685, 553.

\bibitem[]{w05}
Wright, J. T. 2005. PASP, 117, 657.

\end{thebibliography}
\end{document}